\documentclass[12pt]{article} 
\usepackage{url,hyperref,lineno,microtype,subcaption}
\usepackage[onehalfspacing]
{setspace}
\usepackage{pdflscape}
\usepackage{adjustbox}
\usepackage{tabularx}
\usepackage[round,authoryear]{natbib}
\usepackage[utf8]{inputenc}
\usepackage[T1]{fontenc}
\usepackage{lmodern}
\usepackage[margin=1in]{geometry}

\title{A Modular Theory of Subjective Consciousness for Natural and Artificial Minds}

\author{Micha\"el Gillon\thanks{F.R.S.-FNRS Research Director, Astrobiology Research Unit,
University of Li\`ege, Belgium. \emph{Correspondence:} \texttt{michael.gillon@uliege.be}}}

\date{} % empty date

\begin{document}
\maketitle

\begin{abstract}
Understanding how subjective experience arises from information processing remains a central challenge in neuroscience, cognitive science, and AI research. The Modular Consciousness Theory (MCT) proposes a biologically grounded and computationally explicit framework in which consciousness is a discrete sequence of Integrated Informational States (IISs). Each IIS is a packet of integrated information ged with a multidimensional density vector that quantifies its informational richness. When the vector is null, the system remains unconscious; when non-null, its magnitude correlates with the intensity of subjectivity. The higher the information-density, the more an informational packet shapes memory, action, and the continuity of experience.

Information from the body and environment is first adaptively filtered, then processed by specialized modules (abstraction, narration, evaluation, self-evaluation), and integrated into an IIS. The resulting packet, tagged with its information-density vector, is transmitted to behavioral readiness, memory, and decision-making modules, with feedback closing the loop. This architecture explains why states associated with strong subjective intensity exert greater influence on behavior and long-term memory.

MCT thus specifies a functional computational pipeline producing discrete informational units with quantifiable internal structure. Unlike Global Workspace Theory (GWT), Integrated Information Theory (IIT), or Higher-Order Thought (HOT), it treats consciousness not as access, integration, or representation alone, but as the generation of informational packets whose internal density correlates with subjective intensity. By reframing subjectivity as a correlate of an internal information-density tagging signal with functional consequences, MCT bridges clinical observation, cognitive neuroscience, and computational modeling. The model generates testable predictions — such as enhanced memory encoding under stress — and provides a naturalistic blueprint for both biological and artificial systems. In this view, consciousness is not an ineffable essence but an evolvable, quantifiable, and constructible feature of complex information processing.
\end{abstract}


\section{Introduction}
Consciousness remains one of the most intriguing and unresolved phenomena in science. Despite decades of progress in neuroscience, cognitive modeling, and philosophy of mind, no theory has yet provided a comprehensive and computationally explicit explanation of how conscious experience arises, what its functional role is, and why it evolved. While many current frameworks agree that consciousness modulates perception, memory, attention, and decision-making, few specify a full information-processing architecture capable of linking subjective experience to both evolutionary advanes and potential artificial implementation.

Prominent theories offer valuable but partial insights, often focusing on specific cognitive functions or neural correlates rather than a unified computational framework capable of explaining both biological and artificial consciousness. The Global Workspace Theory (GWT) describes the global diffusion of selected information to specialized subsystems \citep{baars1997theater,dehaene2011consciousness}, but leaves unspecified what happens downstream and remains agnostic about subjectivity. Integrated Information Theory (IIT) defines a mathematical measure of consciousness \citep{tononi2004information,oizumi2014phenomenology}, but does not specify the mechanisms by which conscious states are produced, nor how they causally influence behavior in adaptive terms. Higher-Order Theories (HOT) \citep{rosenthal2005consciousness,lau2011hot}, predictive coding frameworks \citep{friston2009free}, and recurrent processing models \citep{lamme2006towards} provide important perspectives, yet do not converge on a unified functional architecture that explains how subjective experience is generated, what computational role it plays, and how it could be instantiated in non-biological systems.

These limitations become salient in relation to the “hard problem” \citep{chalmers1995facing}: why and how do physical processes give rise to subjective experience? Most theories either treat subjectivity as a non-functional byproduct of neural complexity, or as an irreducible property without explanatory power. Neither approach explains how subjectivity might enhance survival by improving memory encoding, behavior prioritization, or adaptive decisions under uncertainty.

In this context, we propose the Modular Consciousness Theory (MCT), a biologically grounded yet substrate-agnostic functional architecture. MCT conceptualizes consciousness as a sequential flow of discrete informational packets — temporally bounded, structured units — generated by modular processing of filtered sensory, interoceptive, and mnemonic inputs. Each unit is a temporally bounded representational state ged with an information-density vector whose magnitude is hypothesized to correlate with the degree of felt subjectivity. This vector biases memory encoding, behavioral readiness, and downstream decision-making. This framing avoids localizationism: modules are defined as computational roles, not rigid brain regions. The model traces a plausible evolutionary trajectory for the gradual addition and integration of these modules, and outlines how similar architectures could be engineered in artificial systems. By treating subjectivity as correlated with an information-density tagging signal, MCT offers falsifiable predictions and a new lens for understanding psychiatric disorders as disruptions of modular integration.

The sections that follow present the structure and logic of MCT. Section 2 introduces the core architecture and information flow through conscious and subconscious modules. Section 3 outlines possible neural correlates. Section 4 compares MCT with existing theories. Section 5 considers implications for synthetic systems. Section 6 discusses empirical predictions, and Section 7 examines psychiatric disorders as cases of modular dysfunction.

\section{Model Description and Evolutionary Aspects}\label{mct}

\begin{figure}[ht]
    \centering
    \includegraphics[width=\textwidth]{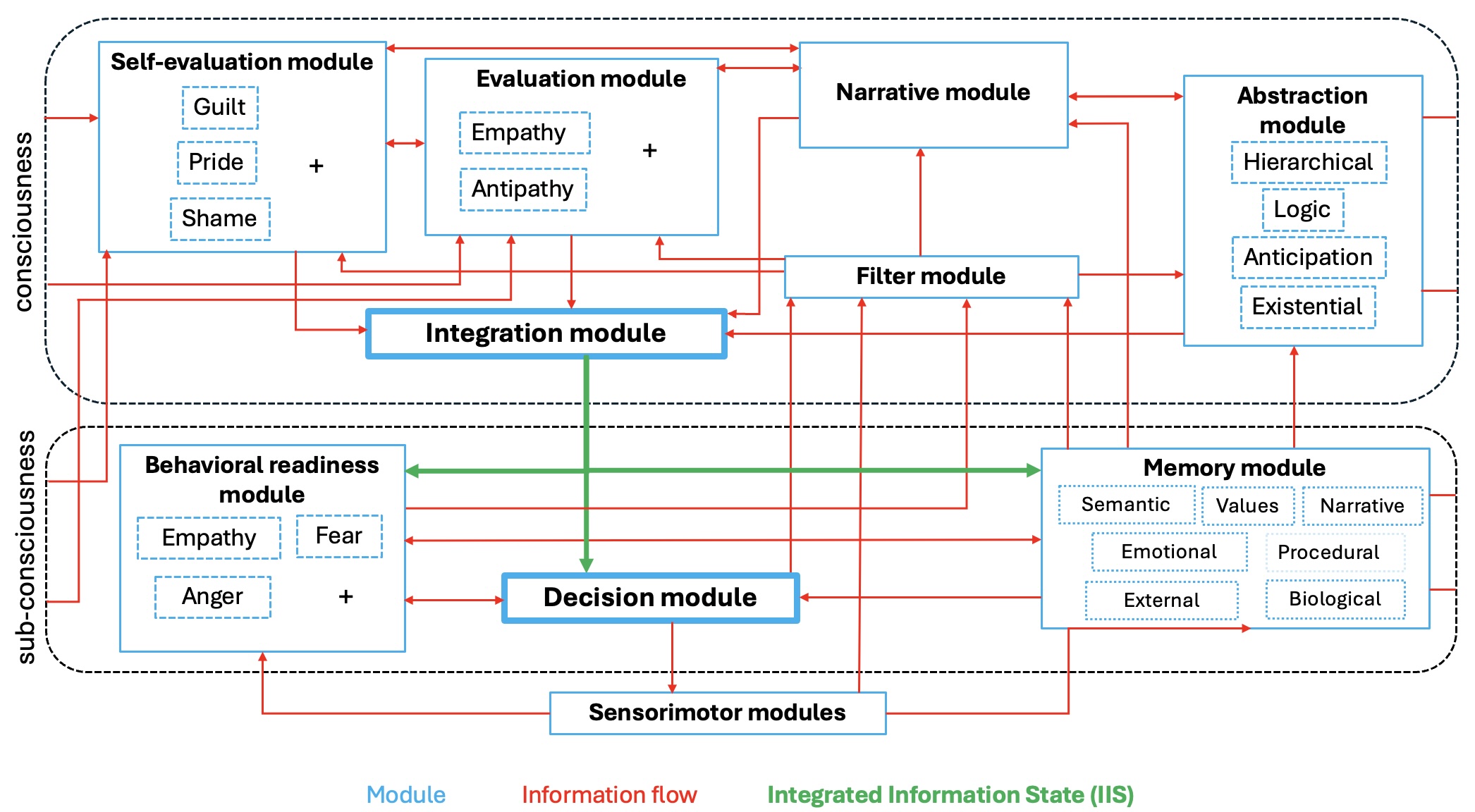}
    \caption{\textbf{Architecture of advanced consciousness in MCT.}  
The full architecture includes evaluation and self-evaluation modules, together with an existential abstraction submodule. Each IIS is generated by the integration module and assigned a multidimensional informational-density vector. These enriched states are propagated to memory, behavioral readiness, and decision modules. This configuration supports introspection, moral reasoning, social cognition, and the construction of a temporally extended self-model. Consciousness is thereby not restricted to the present but spans complex autobiographical representations anchored in narrative and integrative coherence.}
    \label{fig1}
\end{figure}

The Modular Consciousness Theory (MCT) describes cognition as a distributed architecture composed of specialized information-processing modules (Figure~\ref{fig1}). Each module performs a distinct function — such as filtering, abstraction, evaluation, or narrative construction — and operates according to its own computational principles. These modules are defined as functional entities, not as spatially localized brain regions: their physical substrate may involve distributed and anatomically distinct neural components that cooperate to perform a specific computational role. This modular functional organization, supported by neuropsychological, anatomical, and evolutionary evidence \citep{fodor1983modularity, kanwisher2010functional}, underlies key domains of cognition that include perception, motor control, affective regulation, memory, reasoning, and self-reflection. 

In this framework, consciousness does not emerge from a monolithic mechanism, but from a specific form of dynamic coordination: the integration of selected module outputs into a discrete, unified and temporally bounded construct called the \emph{Integrated Informational State} (IIS). Each IIS constitutes the informational content of conscious awareness. It serves as a central hub for memory encoding and the modulation of subsequent behavior.

MCT adopts a functional definition of an \emph{emotion}: a systemic reaction triggered by internal or external stimuli — such as a change in heart rate or muscular tension in response to threat. The informational correlate of that emotion — the signal that can, depending on context, enter conscious awareness and influence memory or behavior — is defined as a \emph{sensation}.

MCT does not seek to define the ontological nature of \emph{subjectivity}, nor to solve the hard problem directly. Instead, it proposes a functional and physically grounded correlate: the subjective feeling arises from an internal process — the system's evaluation of the IIS it has just constructed.

The evaluation of the IIS involves measuring its information density along selected dimensions — such as narrative consistency, emotional salience, autobiographical anchoring, and temporal coherence — which depend on the system's architecture and implementation. The result is a multidimensional vector written as metadata within the IIS, referred to as the \emph{information-density vector} (or \emph{subjectivity vector}).

Although MCT remains agnostic about the ultimate nature of subjective experience, it empirically correlates the intensity of the felt subjectivity with the amplitude of this vector. This allows subjectivity to be treated as an internal signal — a \emph{sensation} in MCT terms — that modulates the likelihood of memory encoding, influences behavioral readiness and decision-making, and contributes to the construction of a coherent stream of conscious experience.

High-amplitude information-density vectors are associated with IISs that are more likely to be stored in memory, influence behavioral readiness and decisions, and support a sense of agency and temporal continuity. Conversely, low-amplitude vectors — such as in sedation, dissociation, or stress overload — correspond to weakly integrated IISs, degraded memory encoding, and diminished behavioral influence.

MCT thus frames consciousness as a graded process: a succession of informational states, each ged with a dynamic information-density vector that determines its cognitive and behavioral weight.

To structure this process, MCT distinguishes three broad categories of modules (Figure~\ref{fig1}):
\begin{itemize}
  \item \textit{Sensorimotor modules}, which govern perception, motor execution, and physiological regulation. These include both somatic and autonomic processes. Although they operate outside of conscious awareness, they provide essential inputs to higher cognitive layers.
  \item \textit{Subconscious modules}, which generate reactive, associative, or automated outputs without invoking subjective awareness. These include mechanisms for memory storage and retrieval, behavioral readiness adaptation, and decisions.
  \item \textit{Conscious modules}, which contribute directly to the construction of the IIS. Their coordinated outputs — abstraction, narration, evaluation, and self-evaluation — form the core content of conscious experience.
\end{itemize}

By layering these components into a flexible and evolvable architecture, MCT outlines a plausible trajectory from reflexive organisms to agents capable of introspection, self-modeling, and moral cognition. Consciousness, in this model, is not a binary state but a graded continuum shaped by the progressive integration of specialized modules.

The following subsections describe how these modules interact in three increasingly complex configurations, each corresponding to a major se in the evolutionary development of consciousness — from basic integration to advanced forms of social cognition.

\subsection{Unconscious Architecture}

\begin{figure}[ht]
    \centering
    \includegraphics[width=\textwidth]{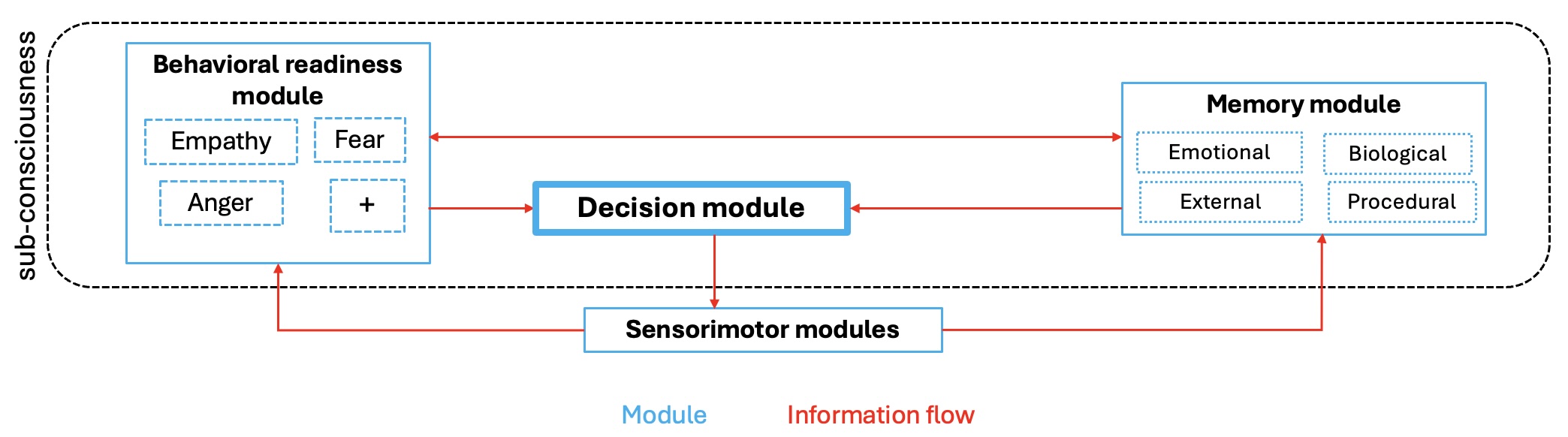}
    \caption{\textbf{Architecture of unconscious agents in MCT.}  
In unconscious architectures, information flows from sensorimotor modules through behavioral readiness and decision systems, with memory restricted to procedural and associative learning. No IIS is generated and thus no subjective signal arises. Behavior remains reactive and stimulus-driven, without abstraction, narrative construction, or integrated self-models.}
\label{fig2}
\end{figure}

The most elementary cognitive configuration described by MCT corresponds to a non-conscious system composed of sensorimotor, behavioral readiness (BR), and memory-related modules, coordinated by a distributed decision module (see Figure~\ref{fig2}). This architecture does not generate IISs and thus lacks informational unification, temporal ging, and any form of self-representation. It supports reflexive behavior, adaptive responses, and basic learning, but operates without an internal model of the system as a unified agent — there is no subjectivity, no narrative, and no awareness of internal states \citep{edelman2000universe, merker2007consciousness}.

In MCT, what are traditionally called \emph{primary emotions} — such as fear, anger, or joy — are redefined functionally as structured states of \emph{behavioral readiness}. These are evolved physiological and motor programs triggered by biologically relevant stimuli and optimized for adaptive response. For instance, the perception of a looming object may activate an escape-oriented BR state, functionally equivalent to what psychology labels as "fear."

A more rudimentary configuration exists in some basal organisms, such as cnidarians. In these systems, local sensorimotor modules for perception, motor control, and homeostatic regulation operate in relative autonomy. These modules may be loosely coordinated by a diffuse neural network acting as a minimal decision center, capable of producing globally organized behaviors (e.g., phototaxis or chemotaxis) \citep{moroz2009independent}. However, such systems lack centralized memory integration and behavioral arbitration \citep{ginsburg2019evolution}.

In the standard unconscious architecture, the \emph{behavioral readiness module} consists of multiple pattern-recognition submodules, each tuned to detect specific configurations of internal or external stimuli. When a match is detected — such as pain, imbalance, or threat — the corresponding submodule emits a high-priority signal proposing a stereotyped physiological and motor response (e.g., withdrawal, freezing, aggression). These signals are transmitted to the \emph{decision module}, which selects among them through mutual inhibition, fixed thresholds, and reinforcement-based heuristics. It does not generate novel behaviors, but acts as a selector, ensuring coherent responses based on priority.

The \emph{memory module} enables plasticity without consciousness. It adjusts the sensitivity and priority of BR submodules based on reinforcement history. Emotional memory biases stimulus-response pathways according to past affective outcomes, while procedural memory modifies behavioral weights based on success or failure. This results in non-symbolic, reflex-like learning — driven not by internal simulation, but by dynamic reweighting of structural associations.

In this architecture, association arises not from inference, but from implicit structural adjustment. The memory system acts as a self-organizing classifier: it strengthens links between co-occurring stimuli and responses, and weakens those associated with failure. Similar paradigms have been explored in artificial agents — from reinforcement learning models to memory-augmented neural networks \citep{graves2016hybrid, miconi2018differentiable}. Unlike in these artificial systems, however, biological memory operates without explicit modeling or symbolic abstraction.

Despite the absence of global integration, unconscious architectures can exhibit considerable behavioral sophistication. Many invertebrates (e.g., insects, arachnids, annelids) and basal vertebrates function within such architectures. While they demonstrate learning and context-sensitive behavior, they show no evidence of internal narrative, volitional inhibition, or mental simulation — features requiring conscious-level integration. Borderline cases exist: some fish and amphibians exhibit behaviors suggestive of abstraction or social learning \citep{ginsburg2019evolution}, potentially reflecting partial recruitment of conscious modules.

In summary, MCT's unconscious architecture provides a functional account of adaptive behavior in the absence of informational integration. It applies to a broad spectrum of biological and artificial systems, and forms the evolutionary substrate upon which more complex, conscious architectures are built.

\subsection{Minimal Conscious Architecture}

\begin{figure}[ht]
    \centering
    \includegraphics[width=\textwidth]{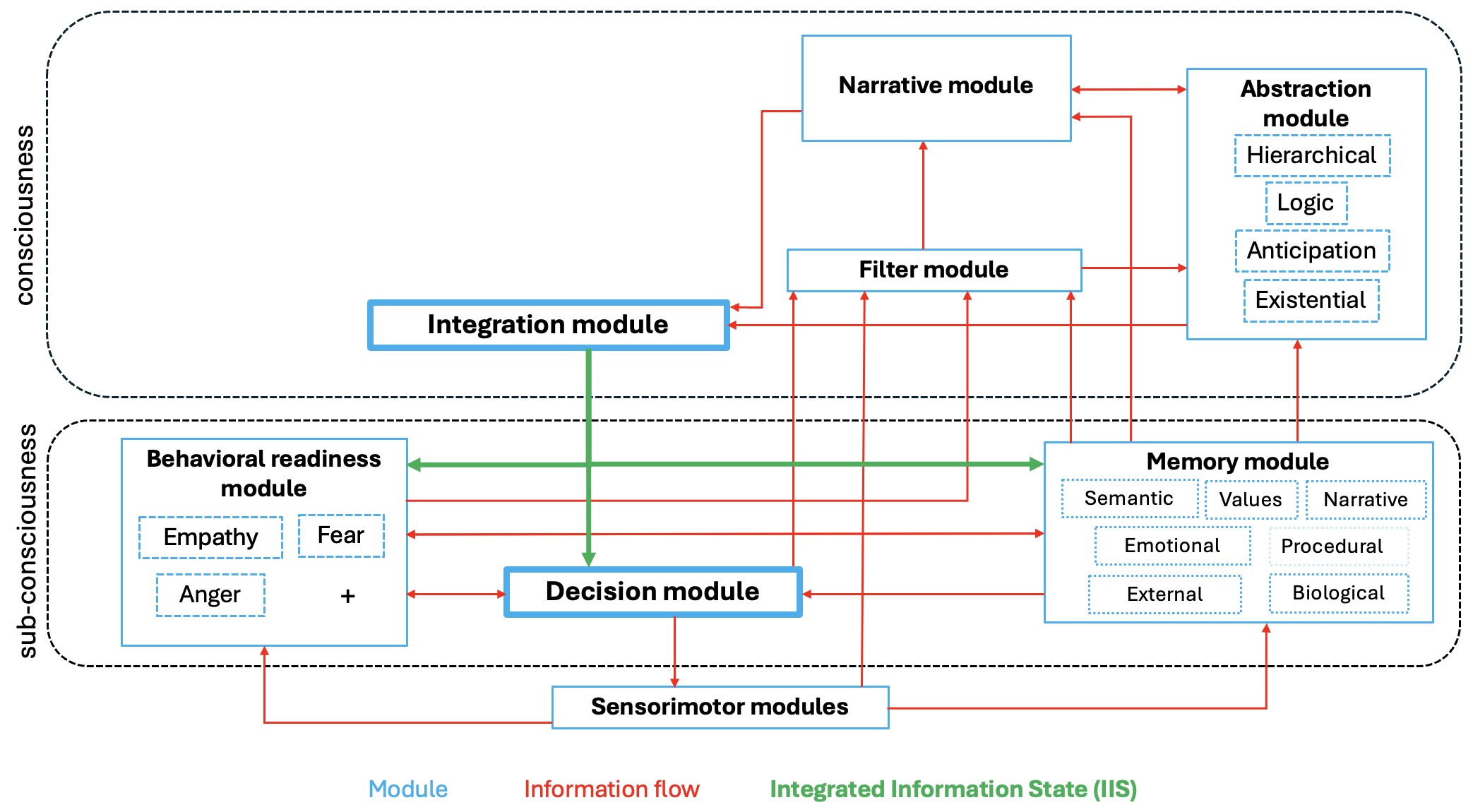}
    \caption{\textbf{Architecture of minimal consciousness in MCT.} 
Subjective consciousness emerges with the addition of abstraction, narrative construction, and a centralized integration module. The resulting IISs are propagated to memory, behavioral readiness, and decision systems, enabling flexible, context-sensitive behavior. This configuration, however, lacks evaluation and self-evaluation modules and thus cannot support moral reasoning, higher-order self-assessment, or socially contextualized cognition.}
   \label{fig3}
\end{figure}

Minimal consciousness, in the MCT framework, emerges with the addition of four key modules: information filtering, narration, abstraction, and integration. Together, they enable the construction of IISs — unified, temporally bounded informational states — which form the basis of conscious-level processing. This configuration does not yet include evaluative or metacognitive modules, and therefore cannot support introspection, moral reasoning, or advanced social cognition (see Figure~\ref{fig3}).

The \textit{filter module} selects a subset of sensorimotor, memory-related, and decision-relevant signals from the subconscious stream and forwards them to the conscious workspace. This selection is context-sensitive — modulated by attentional priorities and feedback from memory — but not based on explicit evaluation. Its function is to regulate bandwidth and enable higher-order modules to operate on a shared, coherent informational substrate \citep{corbetta2002control}. Once filtered, signals are sent in parallel to the abstraction and narration modules, allowing for synchronized processing.

The \textit{narrative module} organizes inputs into a temporally ordered storyline. It maintains continuity across processing cycles and supports the construction of a minimal self-model, anchoring current information to past events \citep{gazzaniga2011who}.

The \textit{abstraction module} enables generalization, reasoning, and prediction through three key functions:
\begin{enumerate}
\item \textit{Hierarchical abstraction} — generalizing across similar situations;
\item \textit{Logical abstraction} — ensuring internal coherence;
\item \textit{Anticipatory abstraction} — simulating possible outcomes.
\end{enumerate}
These functions allow for inhibition of reflexes, basic strategy formation, and context-sensitive planning. Memory serves as the substrate for retrieving patterns and evaluating outcomes.

At the core of this configuration lies the \textit{integration module}, responsible for fusing the outputs of narration and abstraction into a unified IIS. This IIS captures the system's current high-level state, integrating relevant inputs over a short temporal window.

Following its construction, the IIS is analyzed internally through a metacognitive process that estimates its informational density along selected dimensions — such as coherence, relevance, and emotional salience. The result of this analysis is a multidimensional vector that is inscribed as metadata within the IIS. This vector serves as a functional  guiding downstream processes such as memory prioritization, decision influence, and behavioral modulation.

Each IIS — ged with its information density vector — is transmitted to the memory, behavioral readiness (BR), and decision modules. Memory uses the vector to prioritize storage and consolidation. The decision module integrates the IIS into current policy arbitration. The BR module can react to IISs representing not only real-time stimuli but also simulated or remembered scenarios — enabling behavioral responses to internally generated representations.

This architecture thus enables reflex inhibition, flexible learning, goal setting, and rudimentary planning. While lacking introspective depth or self-evaluation, it supports a stable stream of internally structured, temporally coherent informational states. Compared to more advanced systems, the representational and phenomenological richness remains limited, but this configuration marks a functional threshold: from purely reactive behavior to representational guidance.

Such an architecture may be instantiated in organisms such as teleost fish, amphibians, or primitive reptiles, which exhibit goal-directed behavior, learning from context, and limited scenario simulation — without requiring self-reflection or complex social modeling. These systems likely support a minimal narrative self: not introspective, but capable of operating over structured internal representations \citep{ginsburg2019evolution}.
 
\subsection{Advanced conscious architecture}

The advanced conscious architecture described by MCT — illustrated in Figure~\ref{fig1} — builds upon the minimal configuration by incorporating additional high-level modules that enable introspection, moral reasoning, and socially contextualized cognition. In particular, it adds a \textit{self-evaluation module}, an \textit{evaluation module}, and an \textit{existential submodule} within the abstraction system. Together, these support deeper forms of selfhood, symbolic processing, and social understanding.

The \textit{self-evaluation module} compares the current self-representation with internalized values, ideals, and behavioral norms. It generates secondary emotions such as guilt, shame, pride, or self-disgust, each associated with a specialized subcircuit. These emotions presuppose autobiographical continuity, abstract judgment, and a stable narrative self as constructed by the narrative and abstraction modules \citep{tangney2007self}. They allow behavior to be regulated in accordance with ethical principles, social expectations, and long-term identity maintenance \citep{moll2005moral}.

The \textit{evaluation module} enables the attribution of mental states to others — supporting theory of mind, cognitive empathy, and intention modeling. It interacts with memory, abstraction, and narrative processes to simulate the perspectives of others and evaluate their motivations. This capacity enables moral inferences (e.g., “she lied to protect someone”) and context-dependent social behavior. Dysfunctions in this system are associated with impairments in social cognition, moral reasoning, and empathy \citep{frith2006social, blair2005neurocognitive}.

Within the abstraction system, the \textit{existential submodule} allows the construction of symbolic concepts that transcend immediate perception — such as justice, mortality, or destiny. These abstractions support the formation of ideologies, moral codes, and existential narratives, enriching the structural depth and emotional complexity of conscious life \citep{davisson2017social,sedikides2023self}. They underlie uniquely human experiences such as grief, transcendence, or the search for meaning \citep{king2021meaning,guldin2023integrated}.

A further refinement is the emergence of a \textit{linguistic loop}. The linguistic submodule — part of the sensorimotor architecture — executes motor commands for speech and subvocalization. Commands issued by the decision module are routed to this submodule, which produces either overt speech or inner speech. Though not shown individually in Figure~\ref{fig1}, this loop plays a central role in recursive thought. When an IIS triggers internal speech, it initiates a motor intention that is re-perceived and reintegrated into the next IIS. This recursive speech loop reinforces the illusion of a central, continuous “self” — even though no single module holds such identity. The apparent unity of agency — the “I” who thinks or speaks — arises from this recursive integration, not from a central executive \citep{morin2005inner,alderson2023inner}.

This advanced architecture enables a narrative self enriched by introspection, symbolic thought, and cultural embedding. It supports not only flexible planning and adaptive behavior, but also identity maintenance, moral coherence, and existential positioning. The decision module integrates rich IISs with somatic and emotional signals to produce socially situated, temporally extended behavior.

While most fully realized in \textit{Homo sapiens}, similar configurations may exist in less integrated forms in great apes, elephants, or cetaceans \citep{dewaal2016mama, plotnik2006mirror, marino2007cetaceans}. Though these species lack fully developed linguistic loops, their complex social cognition suggests partial implementations of evaluative and introspective processes. In humans, language, education, and cultural transmission consolidate and extend the architecture across generations.

This configuration represents the current apex of the MCT evolutionary trajectory — though further expansions remain conceivable. Each additional module expands the system’s expressive, regulatory, and ethical capacities,  enabling not only greater cognitive flexibility, but also deeper symbolic integration and moral reasoning. In this architecture, subjective consciousness reaches its most developed form: not only capable of experience, but also of reflecting on that experience and deriving meaning from it.

\subsection{The conscious cycle in MCT}

Each conscious cycle begins when the filter module evaluates incoming sensorimotor, emotional, mnemonic, and decision-related signals from the subconscious stream. Access to consciousness is not automatic: each signal is assigned a dynamic salience score, and only those exceeding a context-sensitive threshold are selected. This threshold is adaptive — rising in response to repetition (e.g., habituation) and falling in response to novelty, uncertainty, or goal relevance. Consequently, constant or predictable signals are filtered out over time, allowing the system to prioritize meaningful change and allocate conscious resources efficiently.

Once selected, these signals are simultaneously broadcast to the abstraction, narration, evaluation, and self-evaluation modules. These modules operate on a shared informational substrate, but not in isolation. Their operations are transversal and interdependent: the abstraction module may consult narrative continuity to construct plausible simulations; the evaluation module may rely on narrative structure to infer others’ intentions; the self-evaluation module may invoke abstract norms or imagined perspectives to assess the current self-state.

This horizontal communication — supported by reciprocal signaling pathways — enables each module to refine its output based on the outputs of the others, producing a co-modulated ensemble of interpretations. 

The outputs of these parallel processing streams are subsequently integrated by the integration module into a unified structure: the IIS. The resulting IIS is immediately analyzed. Its informational density is measured along several predefined dimensions, and the resulting values are encoded directly within the IIS as a multidimensional vector. It encodes the perceived amplitude, coherence, salience, and self-relevance of the IIS. 

This IIS is then transmitted to the memory system, where the information-density vector modulates storage priority; to the behavioral readiness module, which can now also react not only to direct stimuli, but to simulated, remembered, or anticipated scenarios; and to the decision module, which interprets the full IIS to guide behavior.

This recurrent loop — salience-based selection, parallel conscious processing, integration, information density ging, and transmission — constitutes a full cycle of consciousness. These cycles unfold rhythmically, at a rate compatible with perceptual update frequencies (typically between 15 and 70 Hz, depending on modality and attention \citep{vanrullen2003perception}), forming the stream of consciousness: a succession of annotated informational states, each carrying its own degree of memory and behavioral relevance. The phenomenal continuity of the conscious awareness emerges from the rapid succession of these discrete states, each corresponding to a fully processed and information-density ged IIS.

This full conscious loop — from filtered input to integrated information-density ged output — corresponds functionally to the high-level architecture illustrated in Figure~\ref{fig1}.

\section{Neurobiological Grounding}\label{neural}

Although MCT is formulated as a functional architecture, its proposed modules can be tentatively associated with large-scale neural systems described in the literature. These associations are speculative and meant to serve as working hypotheses, open to empirical verification and revision. MCT emphasizes functional rather than anatomical modularity: brain regions may contribute to multiple modules depending on context and dynamic recruitment, in line with current network-based views of brain organization \citep{noonan2018socialnetworks, bassett2017network}.

The \textit{narrative module} may correspond to components of the default mode network, including the medial prefrontal cortex, precuneus, and posterior cingulate cortex — regions commonly implicated in autobiographical memory, self-referential processing, and temporal coherence \citep{buckner2008brain,andrewshanna2010functional}. The \textit{abstraction module} could involve lateral prefrontal and parietal areas, often associated with symbolic reasoning, hierarchical structuring, and mental simulation \citep{vincent2008frontoparietal, spreng2010default}.

The \textit{behavioral readiness module} is likely to involve the amygdala, insula, and other limbic structures \citep{phan2002functional}. The \textit{self-evaluation module} — that produces secondary emotions such as guilt or pride — could be linked to activity in the anterior cingulate and ventromedial prefrontal cortices \citep{hao2021emotion}. The \textit{evaluation module}, responsible for cognitive empathy and theory of mind, may involve the temporoparietal junction, medial prefrontal cortex, and precuneus — key components of the social cognition network \citep{frith2006social}.

The \textit{memory module}, which underlies encoding and retrieval, plausibly engages the hippocampus, parahippocampal gyrus, and medial temporal lobe \citep{schacter2007constructive}. The \textit{filter module} — mediating access to conscious processing — might align with frontoparietal attention networks and thalamic relay structures \citep{corbetta2002control}. The \textit{decision module}, which integrates information into action, could involve dorsolateral prefrontal regions, the anterior cingulate cortex, and striatal circuits involved in value-based choice and action selection \citep{miller2001integrative}.

The \textit{integration module} — central to the formation of IISs  — might be instantiated through large-scale synchronization across frontoparietal and midline structures, as suggested by empirical studies of conscious access \citep{dehaene2011consciousness, boly2017consciousness, mashour2020consciousness}. \textit{The information-density tagging} — the measurement of the multidimensional informational density of an IIS — could plausibly involve interactions between the anterior insula, the anterior cingulate cortex, and midline prefrontal regions. These areas are known to participate in salience detection, uncertainty monitoring, and self-referential integration, and may collectively support the computation of relevance or “signal strength” that informs memory encoding, emotional resonance, and decision prioritization  \citep{seeley2007dissociable, menon2015salience}. The intralaminar nuclei of the thalamus, particularly the central lateral nucleus, may play a modulatory role by orchestrating cortical coupling during conscious states \citep{schiff2007behavioral, redinbaugh2020thalamus}.

Importantly, MCT is a data-processing architecture and does not posit a fixed neuroanatomical map. Instead, it proposes that conscious episodes emerge from dynamic interactions between functional modules, each potentially supported by overlapping and distributed networks. The transient synchronization of these networks — particularly in gamma and beta bands — has been proposed as a neural mechanism for binding information into unified conscious episodes \citep{engel2001dynamic, varela2001brainweb}.

Future research may clarify how disruptions to specific circuits affect the structure of conscious experience as predicted by MCT. Interventions such as lesion mapping, transcranial magnetic stimulation (TMS), or transcranial direct current stimulation (tDCS), combined with behavioral and phenomenological assessments, may help determine the causal role of candidate neural substrates in shaping narrative, evaluative, and affective components of consciousness.

The hypothesized neural substrates corresponding to each MCT module are summarized in Table~\ref{tab:mct_neuro}. These associations remain provisional and are intended to guide future empirical testing rather than assert definitive anatomical mappings.

\section{Relationship to Existing Theories}
\label{othertheories}

MCT offers a comprehensive and functionally grounded architecture of consciousness. Unlike most existing models, which emphasize partial aspects — such as GWT, IIT, or HOT — MCT specifies a complete processing pipeline. Information from the body and environment is filtered, transformed by specialized modules, integrated into a discrete informational unit (IIS), and ged with a density vector that determines its impact on memory, behavioral readiness, and decision-making. This explicit sequence connects conscious processing to adaptive behavior in a way that is absent from other frameworks.

A defining feature of MCT is that consciousness is not continuous but discretized into a flux of IISs. The sampling rate of consciousness is thus set by the system’s integration cycle rather than any fundamental physical timescale. Each IIS may be unconscious (vector null) or conscious (vector non-null), making consciousness both discrete and quantifiable.

Crucially, the magnitude of the density vector is hypothesized to correlate with the intensity of subjectivity. This internal information-density ging signal has functional consequences: it modulates the likelihood of an IIS being stored in long-term memory, biases behavioral shifts through primary emotional systems, and influences decision-making. Subjectivity is therefore not treated as an epiphenomenal by-product, but as a measurable correlate of an internal informational property with adaptive value.

In what follows, we compare MCT to leading theories such as GWT, IIT, higher-order and recurrent theories, and predictive coding models, highlighting both points of convergence and the novel aspects that distinguish MCT from all existing accounts.

\subsection{Global Workspace Theory (GWT)}

GWT \citep{baars1997theater, dehaene2011consciousness} proposes that consciousness arises when information is distributed globally to specialized modules in the brain. This idea has inspired several cognitive models in artificial intelligence, such as LIDA \citep{franklin2006lida} and ACT-R \citep{anderson2004integrated}, and aligns with experimental findings showing sudden and widespread cortical activation — often referred to as ``global ignition'' — in response to consciously perceived stimuli.

MCT associates this phenomenon to the initial transfer of information from the filter module to the conscious modules, marking the onset of conscious processing. The widespread activation observed in ignition studies corresponds, in MCT, to the recruitment of conscious modules — including abstraction, narration, integration, evaluation, and self-evaluation — and reflects the early phase of IIS construction. However, MCT introduces a key distinction: this broadcast is not sufficient for consciousness: it merely initiates the integrative process that culminates in the formation of an IIS. In this view, conscious experience emerges not from broadcasting alone, but from the construction of a structured informational state ged with information density.

MCT explains why emotionally charged or autobiographically relevant stimuli reach awareness more reliably: the filter module prioritizes inputs based on their intensity, emotional valence, and mnemonic relevance, enabling them to cross the threshold for conscious access. Unlike GWT, which remains agnostic about the nature and function of subjectivity, MCT posits an internal neurofunctional signal whose magnitude correlates with reported subjective intensity and is embedded in each IIS as a graded metacognitive marker. This marker reflects coherence, personal relevance, and informational richness, and plays a critical role in memory consolidation, behavioral readiness, and decision-making. In this sense, MCT extends GWT by specifying both the internal architecture and the computational role of subjectivity within a biologically plausible system.

\subsection{Integrated Information Theory (IIT)}

IIT \citep{tononi2004information, oizumi2014phenomenology} offers a mathematical framework for quantifying consciousness via $\Phi$, a scalar measure of integrated information. Empirical support for IIT includes the Perturbational Complexity Index (PCI), which correlates with changes in consciousness under anesthesia, sleep, or brain injury \citep{casali2013theoretically}.

MCT offers a functional reinterpretation of these findings. It does not treat $\Phi$ or PCI as direct measures of consciousness, but as proxies for the brain’s capacity to generate rich informational structures — what MCT calls IISs. In conscious states, the system produces a stream of IISs that vary in subjective intensity, depending on their coherence, autobiographical relevance, and emotional salience. Under sedation or fatigue, the formation of such IISs becomes sparse and poorly ged, reducing subjective vividness and mnemonic encoding — though not necessarily abolishing consciousness altogether.

This perspective anchors IIT’s formalism within a modular cognitive architecture. In MCT, $\Phi$ is reconceptualized as a global indicator of the system’s ability to coordinate its conscious modules — abstraction, narration, evaluation, etc. — into integrated, self-relevant informational episodes. In this sense, $\Phi$ may correlate with the potential for strong information-density ging, but the  itself — the metacognitive evaluation of the IIS — requires additional processing not addressed by IIT.

Unlike IIT, which equates high $\Phi$ with consciousness itself, MCT draws a crucial distinction: integrated information is necessary but not sufficient. Conscious experience requires not only the construction of an IIS, but also its evaluation — a functional step in which the system assigns a graded information-density vector. This vector encodes not just integration, but significance. It is this signal, and not $\Phi$ alone, that governs memory encoding, emotional regulation, and decision-making.

Thus, MCT retains the mathematical insights of IIT but embeds them in a biologically and functionally grounded model — one capable of explaining not only how information is integrated, but how it becomes experienced.

\subsection{Higher-Order and Recurrent Models}

Higher-Order Theories (HOT) \citep{rosenthal2005consciousness} and recurrent processing models \citep{lamme2006towards} propose that consciousness arises either when a mental state becomes the object of a higher-order representation, or when perceptual activity becomes reentrant and globally accessible. These models emphasize reflexive awareness and recursive accessibility, but often lack a detailed account of the minimal structural and functional conditions required for conscious experience.

MCT addresses this by specifying the precise modular architecture needed to generate a self-relevant informational state. In this framework, conscious experience emerges when the outputs of the abstraction, narration, and evaluation modules are jointly integrated into a temporally bounded structure — the IIS. This state is not simply a perceptual broadcast or a metarepresentation, but a unified informational episode enriched by internal coherence, narrative framing, and contextual relevance.

Unlike classical HOTs, which posit explicit metarepresentations or linguistic self-reference, MCT embeds reflexivity within the structure of the IIS itself. The integration of abstract inferences, narrative continuity, and evaluative signals produces a state that is both representational and self-relevant. The system’s internal response to this configuration is formalized as an information-density vector — a multidimensional signal encoding coherence, salience, and self-significance.

Recurrent processing, from the MCT perspective, is interpreted not as the essence of consciousness, but as a possible neural mechanism supporting the sustained activation, synchronization, and binding of conscious modules. What defines consciousness in MCT is not recurrence per se, but the construction and ging of cohesive IISs — structures that can guide memory, modulate emotion, and inform context-sensitive behavior.

\subsection{Predictive Coding Models}

Predictive coding and free-energy minimization frameworks \citep{friston2009free} offer powerful models of perception and action. These approaches conceptualize the brain as a hierarchical inference engine, continuously generating predictions about incoming sensory data and minimizing prediction errors across levels of representation.

MCT incorporates this inferential mechanism within its abstraction module, which supports scenario simulation, expectation formation, and the construction of temporally structured representations. However, MCT extends beyond predictive coding by specifying the additional modular infrastructure required for conscious experience.

Specifically, MCT outlines how predictive inferences are (1) selectively filtered for conscious access, (2) embedded in narrative structures, (3) evaluated for personal and social relevance, and (4) integrated into IISs whose internal properties are computationally assessed. This assessment — performed by the integration module — generates a multidimensional vector encoding the coherence, stability, and self-relevance of the IIS. This internal signal modulates memory prioritization, behavioral readiness, and decision-making.

Whereas predictive coding explains how organisms anticipate and adapt to sensory input, MCT explains how a subset of these inferences enters consciousness, becomes structured into episodes, and influences long-term behavior. In this sense, MCT provides a functional architecture that transforms inferential processes into subjectively experienced, narratively organized, and behaviorally actionable informational states.

\subsection{Summary}

MCT builds upon the core contributions of leading consciousness theories while integrating them into a unified, modular, and biologically plausible architecture. It incorporates the global accessibility emphasized in GWT, the integration principles of IIT, the recursive organization of HOT, and the inferential structure of predictive coding — but embeds these within a detailed system of interacting modules, each with defined inputs, outputs, and functional roles.

Importantly, MCT reframes experimental markers such as global ignition and the Perturbational Complexity Index not as direct indicators of consciousness, but as signatures of the system’s capacity to construct complex and internally evaluated informational states. These states — the IISs — form the backbone of conscious processing in the model.

By articulating how such states are selected, structured, annotated, and transmitted across memory, decision, and action systems, MCT offers a testable and implementation-ready framework applicable to both biological and artificial agents. Its emphasis on functional organization over abstract phenomenology enables a pragmatic and scalable approach to modeling conscious cognition.

\section{Artificial Consciousness and Synthetic Implementation}\label{IA}

A distinctive strength of MCT lies in its immediate applicability to synthetic systems. Unlike most theories of consciousness — often abstract, metaphysical, or tightly coupled to neurobiology — MCT provides a fully specified functional architecture. It defines the sequence of computations required to process sensory information, evaluate its significance, integrate it with internal models, and guide adaptive behavior.

Some prior implementations of conscious-like architectures in AI — notably those inspired by the Global Workspace Theory (GWT) \citep{franklin2006lida, anderson2004integrated} — have focused on broadcasting information across specialized modules. However, these architectures often lack the structural depth required for coherent episodic representation and behavior modulation. In MCT, consciousness is not broadcasting; it is the full processing loop that filters, evaluates, unifies, and exploits information to steer context-sensitive action.

This loop includes sensory filtering, abstraction, narrative construction, evaluative appraisal, and self-monitoring. Their outputs are fused by the integration module into an IIS — a temporally bounded informational unit annotated with internal information-density values. This structure guides memory encoding, action readiness, and decision-making, while also enabling the simulation of coherent, agent-centered episodes.

In this framework, the impression of being a subject is not a metaphysical property, but a byproduct of representational coherence. The self-model is not an entity but a dynamic construct: an internal reference point used to simplify memory encoding, prediction, and behavioral regulation. The ging of informational states reflects their internal consistency, salience, and relevance to this modeled agent — forming the basis of coherent, goal-directed behavior.

MCT is thus uniquely suited for artificial implementation. Its modular design can be instantiated using standard components in AI systems: pattern recognition, narrative modeling, relevance assessment, and memory management. Each cycle produces an IIS — a structured informational snapshot associated with predictive and behavioral value — forming a continuous stream of annotated episodes.

By framing consciousness as a computational strategy for adaptive behavior optimization, MCT provides a concrete blueprint for building agents capable of contextual modeling, coherent memory formation, and flexible action — whether biological or artificial.

\subsection{From Inputs to Integration: A Full Synthetic Cycle}

In an MCT-based synthetic agent, each conscious cycle begins with raw input data — for example, visual input from a camera array, auditory signals from a microphone, or proprioceptive states from internal sensors. These inputs are processed first by low-level unconscious modules:
\begin{itemize}
    \item \textit{Sensorimotor module}: Converts raw sensory data into abstracted perceptual tokens (e.g., object maps, speech-to-text transcripts, environmental features).
    \item \textit{Behavioral readiness module}: Detects reflex patterns based on threat, reward, or discomfort signatures (e.g., high-pitch vocalizations, rapid motion, internal tension), potentially triggering fast responses or influencing conscious access thresholds.
    \item \textit{Memory module}: Contributes previously encoded data that matches current stimuli via associative retrieval.
\end{itemize}

These data streams converge at the \textit{filter module}, which prioritizes a subset for conscious access based on salience scores. Salience may reflect input intensity, novelty, or autobiographical relevance, modulated via feedback from memory. The selected signals are then broadcast to all conscious information processing modules — \textit{abstraction}, \textit{narration}, \textit{evaluation}, and \textit{self-evaluation} — in parallel \citep{corbetta2002control, engel2001dynamic}.

Each conscious module processes its share of the incoming stream:
\begin{itemize}
    \item \textit{Abstraction module}: Generalizes patterns, simulates possible futures, and updates internal models via a predictive coding or deep reinforcement learning engine \citep{friston2009free}.
    \item \textit{Narrative module}: Arranges temporal content into coherent story-like sequences, using structures akin to transformer-based architectures trained on inner or social dialogue \citep{vaswani2017attention}.
    \item \textit{Evaluation module}: Interprets intent, social cues, or potential outcomes using theory-of-mind algorithms, value hierarchies, or symbolic utility functions.
    \item \textit{Self-evaluation module}: Re-ranks or suppresses outputs based on coherence, contradiction detection, or internalized norms.
\end{itemize}

\subsection{Integration and Internal Information-Density Tagging}

Once module-level processing is complete, results are aggregated by the \textit{integration module}, producing a coherent IIS. Each IIS contains:
\begin{itemize}
    \item Abstracted symbolic content;
    \item Temporally ordered narrative structure;
    \item Evaluative assessments of risk, value, or intent.
\end{itemize}

The IIS is then read and assigned a multidimensional information-density vector. This vector encodes relevance metrics such as novelty, coherence, goal alignment, and urgency. It modulates memory encoding, decision arbitration, and behavioral readiness.

The completed IIS is propagated to three main systems:
\begin{enumerate}
    \item \textit{Memory system}: Stores or discards the IIS depending on information-density level. Dynamic architectures (e.g., Differentiable Neural Computers \citep{graves2016hybrid}) enable high-information-density IISs to be retained and reinforced.
    \item \textit{Decision module}: Selects context-sensitive behavior based on current IIS and information-density vector.
    \item \textit{Behavioral readiness system}: Triggers expressive or preparatory responses as needed.
\end{enumerate}

\subsection{Language, Expression, and Inner Speech}

To enable communication and recursive introspection, an advanced MCT agent could include a \textit{linguistic module}, modeled as a syntax generator (e.g., symbolic parser or transformer-based decoder). This module would convert integrated content into:
\begin{itemize}
    \item \textit{External speech}: Used to express decisions, requests, or affective states.
    \item \textit{Internal speech}: Reintegrated in the next cycle via the filter module, enabling recursive self-modeling and higher-order thought \citep{morin2005inner, alderson2023inner}.
\end{itemize}

This loop reinforces the appearance of a unified “self” and allows symbolic abstractions to influence memory, behavioral readiness, and long-term strategy — completing the minimal architecture for introspective agency.

\subsection{Behavioral and Experimental Predictions}

MCT provides behavioral criteria to test artificial subjectivity. A conscious synthetic agent should:
\begin{itemize}
    \item Preferentially store IISs ged with high subjective relevance;
    \item Modify behavior based not only on sensory input, but on narrative structure and internal salience;
    \item Reactivate past IISs in ambiguous, affectively charged, or decision-critical contexts;
    \item Generate linguistic outputs that reflect internal evaluation and narrative coherence.
\end{itemize}

These predictions can be experimentally tested through memory retention, stress responses, and narrative adaptation tasks. They align with emerging methods in synthetic phenomenology and machine self-modeling \citep{chrisley2009synthetic,graziano2019attention,gamez2018human}.

\section{Empirical Support and Testable Predictions of MCT}\label{testability}

Any scientific theory of consciousness must not only generate testable predictions, but also account for established cognitive and neurophysiological findings. MCT meets both criteria by offering a unified modular architecture that reinterprets diverse empirical results through the lens of IISs.

In this framework, each IIS constitutes a temporally bounded unit of conscious content. Its functional salience — quantified by the norm of its internal information-density vector — reflects the intensity of conscious processing at that moment. This metric enables MCT to formulate specific, testable predictions across three primary domains: memory encoding, decision-making, and behavioral readiness.

\subsection{Memory Consolidation}

MCT predicts that memory retention depends primarily on the informational relevance of the IIS at the moment of encoding, rather than on the objective properties of the stimulus. This prediction aligns with extensive evidence from research on emotional memory, flashbulb phenomena, and stress-related modulation of memory via amygdala–prefrontal circuitry \citep{arnsten2009stress, roozendaal2009amygdala}. Episodic memory studies consistently show enhanced retention for emotionally salient, autobiographically relevant, or narratively coherent content — even when controlling for perceptual complexity \citep{tulving1983elements, cahill2003enhanced, dolcos2004interaction}.

Under MCT, each IIS is associated with an information-density vector that reflects perceived coherence, emotional impact, and personal significance. The amplitude of this vector modulates consolidation likelihood: higher values increase the probability that the IIS will be stored and later retrieved.

This prediction can be tested using paradigms that correlate post-encoding memory performance with either subjective ratings or neural indicators of integrative processing — such as frontoparietal synchronization or evoked complexity metrics. MCT expects a strong positive relationship between integration-driven ging and long-term retention, particularly under conditions of novelty, uncertainty, or emotional engagement.

\subsection{Decision Bias and Behavioral Prioritization}

In MCT, the decision module evaluates the current IIS and uses its associated information-density vector to arbitrate among potential actions. The amplitude of this vector — reflecting informational coherence, emotional intensity, and personal relevance — modulates behavioral urgency. As a result, MCT predicts that agents will preferentially act on information with high internal salience, even when its objective utility is equal or lower.

This prediction is consistent with well-documented cognitive biases, including:
\begin{itemize}
  \item framing effects and moral decision distortions \citep{greene2001fmri, kahneman2011thinking},
  \item action biases linked to autobiographical memory retrieval \citep{conway2000sms},
  \item affect-driven decision-making and somatic marker effects \citep{bechara2000emotion, dunn2006somatic}.
\end{itemize}

Experimental validation could involve decision tasks in which reward probabilities are matched, but internal relevance or narrative coherence varies. MCT expects stronger behavioral responses and shorter latencies for IISs with higher integrative salience — measurable through subjective reports, physiological arousal, or integrative neural signatures.

\subsection{Internal Amplification of Behavioral Readiness}

MCT predicts that shifts in BR can be triggered not only by raw sensory input, but also by internally integrated representations encoded in an IIS. Once a scenario is integrated — even if simulated or retrieved from memory — it can feed back into the BR module and produce physiological or action-oriented states. This feedback loop explains:

\begin{itemize}
  \item action tendencies elicited by imagined or remembered situations \citep{ji2016emotional, montijn2021emotion},
  \item anticipatory activation during future-oriented thinking \citep{peters2010future},
  \item somatic responses evoked by internal narrative reflection \citep{tangney2007self}.
\end{itemize}

MCT therefore predicts that the intensity of BR activation should correlate with the information-density vector of the IIS, even when perceptual input is held constant. This can be tested by comparing physiological responses (e.g., heart rate, skin conductance) across internally generated scenarios with varying degrees of integration, relevance, and coherence.

\subsection{Testable Correlations}

These three predictions lead to specific, quantifiable expectations:
\begin{itemize}
  \item \textit{Memory:} Higher information-density vector amplitude $\Rightarrow$ greater retention and reconsolidation likelihood.
  \item \textit{Decision:} Higher information-density vector amplitude $\Rightarrow$ increased action probability and faster response.
  \item \textit{Behavioral Readiness:} Higher information-density vector amplitude $\Rightarrow$ stronger internal activation, even in the absence of new external input.
\end{itemize}

These effects should be observable under both naturalistic and experimentally controlled conditions. In artificial systems, if such patterns emerge from an architecture implementing MCT principles — including integration, informational ging, and prioritized storage — it would support the emergence of functionally analogous subjective processing.

\subsection{Compatibility with Existing Observations and Pathologies}

In addition to generating testable predictions, MCT provides an integrative framework for reinterpreting a wide range of empirical findings across neuroscience, psychology, and psychiatry. Several well-known phenomena can be understood as resulting from incomplete, degraded, or disjointed formation of IISs.

\begin{itemize}
\item \textit{Blindsight}, \textit{implicit learning}, and \textit{unconscious priming} can be reframed as cases where stimuli influence behavior without giving rise to a coherent, consciously accessed IIS.

\item \textit{Split-brain experiments} illustrate what happens when key conscious modules — such as narration and evaluation — operate in isolation, without mutual integration. This results in fragmented or conflicting conscious outputs.

\item \textit{Anesthesia and deep sleep} correspond to states where IIS formation becomes sparse or severely degraded: the outputs of conscious modules may be reduced or uncoordinated, and subjective s diminish in amplitude, leading to a marked drop in conscious vividness.
\end{itemize}

These reinterpretations highlight how MCT offers a unifying lens for diverse findings typically treated in isolation. Building on this foundation, the next section examines how psychiatric and neurological disorders may result from specific disruptions within the modular architecture proposed by MCT.

\subsection{Revisiting the Libet Experiments with MCT}

In Libet’s paradigm, participants perform a simple action (e.g., finger flexion) at a self-chosen moment \citep{libet1985unconscious}. From an MCT perspective, the instruction forms a weakly tagged IIS — low urgency, minimal autobiographical relevance — which enters the decision module as a soft suggestion. The actual movement is often triggered by unconscious fluctuations in readiness signals \citep{schurger2012accumulator}, with conscious awareness arising retrospectively once sensory consequences are integrated. This explains the delay between neural onset and reported volition.

When the system instead generates a strongly tagged IIS (“move now”), the decision module treats it as a prioritized command, producing deliberate but slower execution due to the integration cost. Reflexive actions bypass this loop, remaining faster but less aligned with agency.

MCT thus reframes free will: volition is not the act of a metaphysical self, but the internal signal of action selection from a strongly tagged IIS. The “self” is an emergent construct organizing integration; freedom becomes a computationally useful illusion rather than an irreducible property.

\section{Clinical Applications: Modularity and Mental Disorders}\label{psy}

MCT provides a modular framework for interpreting psychiatric and neurological conditions in which conscious experience is altered or fragmented. Rather than invoking a unitary dysfunction of ``consciousness,'' it links symptoms to specific failures in modular dynamics: 
\begin{itemize}
  \item deficient \textit{generation} of outputs (e.g., abstraction or evaluation),
  \item faulty \textit{integration} into coherent IISs,
  \item aberrant \textit{modulation} of salience or information-density tagging.
\end{itemize}

Such breakdowns distort subjectivity, autobiographical continuity, affect regulation, and behavioral coherence. MCT thus reinterprets diverse disorders as failures in constructing, weighting, or deploying informational states, offering a principled lens for clinical explanation.

\subsection{Psychopathy}

Psychopathy is a personality disorder characterized by emotional detachment, lack of remorse, and manipulativeness \citep{hare1999without, patrick2018handbook}. Neuroimaging studies consistently implicate anomalies in the amygdala, ventromedial prefrontal cortex (vmPFC), and anterior cingulate cortex — circuits central to affect regulation, moral reasoning, and behavioral inhibition \citep{kiehl2006psychopathology, blair2007amygdala_vmPFC, gao2009neurodevelopmental, koenigs2010neural, motzkin2011reduced}. Within MCT, psychopathy reflects intact abstraction, planning, and memory modules, but dysfunctional evaluative circuits. Empathy-related submodules are underactive, guilt and shame tagging in self-evaluation is hypoactive, and antipathy signals may be amplified. As a result, IISs are generated without inhibitory moral weighting. The narrative system integrates antisocial actions into a coherent storyline, while preserved theory-of-mind capacities enable manipulation decoupled from affective resonance. Psychopathy thus illustrates how consciousness can be cognitively intact yet ethically hollow, when evaluative signals fail to modulate memory, decision-making, and behavior.

\subsection{Malignant Covert Narcissism}

Malignant covert narcissism combines narcissistic vulnerability with traits of manipulation and control \citep{kernberg1984severe,goldner2010malignant,jauk2021neuroscience,schmidt2023structural,somma2024assessing}. Outwardly, individuals may present as hypersensitive or exceptionally empathetic, but their cognitive architecture is organized around the protection of a rigid self-image.

In MCT terms, the self-evaluation module is reorganized: guilt is redirected into self-victimization, shame is projected outward, and pride hypertrophied. The narrative module is enslaved to a fixed schema of the self, such that new experiences are filtered not to maintain autobiographical coherence but to preserve this schema. Memory becomes structurally biased: threatening content suppressed, self-affirming content selectively reinforced \citep{rhodewalt2002narcissus,sen2023mediating}. The result is a flux of IISs consistently tagged to support the same frozen storyline.

Abstraction, particularly its existential subcomponents, remains disengaged from emotional weighting. Concepts such as loss or mortality are understood but lack resonance. Emotional empathy is absent, while cognitive empathy remains intact and is applied instrumentally. Decisions are therefore shaped by IISs whose priority tags are dominated by the urgency to protect the self-schema, not by contextual or relational balance.

This architecture leads to behaviors where others are treated as roles within the individual’s narrative. In extreme cases, it may justify abandoning a partner in crisis, denying shared intimacy, or pursuing institutional and reputational strategies of control — all experienced internally as consistent with the preserved self-image.

MCT characterizes this profile as a form of “narrative capture”: the modules of consciousness remain active, but their outputs are subordinated to the maintenance of a static self-concept. This distinguishes it from psychopathy, where IISs are coherent but emotionally flat. Here, subjectivity is not absent but structurally looped around the defense of identity, producing rigidity, resistance to contradiction, and a capacity for relational harm while maintaining apparent coherence.

\subsection{Severe Dissociative States}

Severe dissociative phenomena — such as those seen in Dissociative Identity Disorder (DID), complex PTSD, or trauma-induced dissociation — reflect a breakdown of conscious integration. In MCT, they correspond to a disruption in the generation of stable IISs, caused by desynchronization between integration, narrative, memory, and behavioral readiness modules.

Neuroimaging studies report hypoactivation of default mode network (DMN) regions and disrupted DMN–limbic connectivity \citep{brand2009dissociative, reinders2012fact, modesti2022functional, taib2023neural}. This aligns with MCT’s prediction that dissociation prevents the integration module from producing IISs enriched with a subjective density vector.

During episodes, sensory and unconscious processing remain partly intact, enabling superficially coherent behavior, but the conscious system fails to generate unified, subjectively tagged IISs. Experience loses introspective presence, autobiographical anchoring, and temporal coherence. In DID, this explains how complex behaviors may occur without later recollection: either no IIS was generated, or it was tagged within an identity-specific subsystem inaccessible to the dominant narrative stream.

Autobiographical memory is fragmented, especially around trauma \citep{robjant2010emerging, thome2020neural}, reflecting failed transfer from integration to memory or disrupted narrative linkage. Dissociation often serves as an adaptive bypass: the system suppresses integration when full IIS generation would overload affective capacity.

MCT thus interprets dissociation not as unconsciousness but as a pseudo-conscious state: perception and behavior persist, yet without introspective awareness or stable identity. Severity reflects the degree of modular desynchronization and the inability to sustain subjectively integrated IISs.

\subsection{Bipolar Disorder}

Bipolar disorder alternates between depressive and manic phases, each with marked shifts in affect, cognition, and behavior \citep{goodwin2007manic}. In MCT terms, it reflects an instability in the calibration of the integration module, producing oscillations in emotional valence, narrative coherence, and subjective tagging of IISs.

During mania, abstraction and narrative systems are hyperactive: expansive simulations and idealized self-concepts dominate, while self-evaluation is dampened. IISs acquire exaggerated density vectors, amplifying salience, autobiographical weight, and euphoria. In depression, by contrast, narrative collapses into punitive ruminations, abstraction becomes rigidly pessimistic, and density vectors flatten or turn aversive.

These poles may arise from neurochemical fluctuations destabilizing integration thresholds and modulatory coupling with abstraction and evaluative circuits. Bipolar disorder thus exemplifies a dysregulation of conscious gain: the amplitude and coherence of IISs oscillate beyond adaptive ranges, requiring restored synchrony between narrative, abstraction, and evaluation.

\subsection{Discussion}

By linking symptoms to specific modular dysfunctions, MCT provides a coherent framework for interpreting disorders of consciousness (Table~\ref{tab1}). Rather than classifying pathologies only by external criteria, it identifies which architectural configurations are disrupted — opening the way to modular diagnostics and interventions.

Across conditions from dissociation to narcissistic and psychopathic profiles, a common principle emerges: the function of consciousness is not to mirror reality, but to construct an internally consistent informational flow optimized for adaptive behavior. In MCT, the flux of IISs is shaped less by raw input than by constraints such as narrative stability, predictive utility, and motivational alignment.

This principle extends to ordinary cognition. Memory distortions, belief perseverance, and post-hoc rationalization reflect the same bias: the system privileges continuity of processing over empirical fidelity. Consciousness is not designed for accuracy, but for maintaining a stable self-in-context.

Malignant covert narcissism offers a stark illustration. Here, the idealized self-image functions as a rigid survival core. The narrative module selectively filters and reorganizes memory to preserve this schema, while the self-evaluation system redirects guilt into self-victimization and externalizes shame. IISs are thus tagged not for balanced integration, but for identity protection. Others become roles in this script, and contradictory evidence is suppressed. This produces a consciousness that is modularly functional yet epistemically distorted — coherent for the individual, destabilizing for others.

MCT therefore suggests that the persistence of identity and informational flow can endure even when accuracy collapses. Consciousness may remain operationally stable while epistemically dysfunctional, a principle relevant both to pathology and everyday cognition.

\section{Conclusion}\label{conclusion}

The Modular Consciousness Theory (MCT) proposes a biologically grounded and computationally explicit model of consciousness. It defines conscious experience not as a continuous and ineffable flow, but as a sequence of discrete informational units — Integrated Informational States (IISs). Each IIS is accompanied by a multidimensional information-density vector that quantifies its internal informational richness and functions as a control signal for memory encoding, behavioral weighting, and decision guidance.

Unlike Global Workspace Theory (GWT), Integrated Information Theory (IIT), or Higher-Order Thought (HOT), which characterize consciousness in terms of access, integration, or higher-order representation, MCT specifies a discrete computational pipeline that transforms filtered inputs into prioritized informational packets. Subjectivity, in this framework, is not a metaphysical essence nor directly measured by the system; rather, the magnitude of the internal information-density vector is hypothesized to correlate with the intensity of reported subjectivity. 

The apparent stability of the self arises from the narrative module, which organizes successive IISs into temporally ordered sequences. Agency and identity thus emerge as constructs required for coherence across informational units, rather than as intrinsic entities.

Because MCT is formulated in computational terms, it applies equally to biological and artificial systems. Any architecture combining filtering, abstraction, integration, narration, and internal information-density ging could, in principle, generate agent-relative awareness. This makes MCT both a naturalistic blueprint for synthetic consciousness and a framework for testing the minimal conditions under which subjectivity emerges.

MCT does not attempt to resolve the ontological nature of feeling. Instead, it reframes the “hard problem” as a measurable hypothesis: the intensity of subjectivity correlates with the amplitude of the information-density vector associated with each IIS. If this correlation proves consistent and predictive, consciousness becomes a tractable feature of information processing. As with life or heat, explanatory progress comes not from invoking irreducible essences but from uncovering the architectures and processes capable of reproducing the full range of observed phenomena — including subjective report.

\noindent

\newpage

\section{Data availability} 
This work is purely theoretical, and no data were used.

\section{Code availability} 
This work is purely theoretical, and no code was used.

\section{Acknowledgments}
M.G. is F.R.S-FNRS Research Director. 

\section{Author Contributions Statement}
M.G. conceived the theory, developed the model, conducted the literature review, wrote the manuscript, and prepared all figures and supplementary materials. No other individuals contributed to the conceptualization, writing, or revision of this work.

\section{Competing Interests Statement}
The author declares no competing interests.

\setlength{\parindent}{0pt}  
\setlength{\parskip}{20pt}    

\bibliography{MCT}

\begin{thebibliography}{94}
\providecommand{\natexlab}[1]{#1}
\providecommand{\url}[1]{\texttt{#1}}
\expandafter\ifx\csname urlstyle\endcsname\relax
  \providecommand{\doi}[1]{doi: #1}\else
  \providecommand{\doi}{doi: \begingroup \urlstyle{rm}\Url}\fi

\bibitem[Alderson-Day and Pearson(2023)]{alderson2023inner}
Ben Alderson-Day and Joel Pearson.
\newblock What can neurodiversity tell us about inner speech, and vice versa? a theoretical perspective.
\newblock \emph{Cortex}, 168:\penalty0 193--202, 2023.
\newblock \doi{10.1016/j.cortex.2023.08.008}.

\bibitem[Anderson(2004)]{anderson2004integrated}
John~R Anderson.
\newblock An integrated theory of the mind.
\newblock \emph{Psychological Review}, 111\penalty0 (4):\penalty0 1036--1060, 2004.
\newblock \doi{10.1037/0033-295X.111.4.1036}.

\bibitem[Andrews-Hanna et~al.(2010)Andrews-Hanna, Reidler, Sepulcre, Poulin, and Buckner]{andrewshanna2010functional}
Jessica~R Andrews-Hanna, Jeffrey~S Reidler, Jorge Sepulcre, Richard Poulin, and Randy~L Buckner.
\newblock Functional-anatomic fractionation of the brain’s default network.
\newblock \emph{Neuron}, 65\penalty0 (4):\penalty0 550--562, 2010.
\newblock \doi{10.1016/j.neuron.2010.02.005}.

\bibitem[Arnsten(2009)]{arnsten2009stress}
Amy~F.T. Arnsten.
\newblock Stress signalling pathways that impair prefrontal cortex structure and function.
\newblock \emph{Nature Reviews Neuroscience}, 10\penalty0 (6):\penalty0 410--422, 2009.
\newblock \doi{10.1038/nrn2648}.

\bibitem[Baars(1997)]{baars1997theater}
Bernard~J Baars.
\newblock In the theater of consciousness: Global workspace theory, a rigorous scientific theory of consciousness.
\newblock \emph{Journal of Consciousness Studies}, 4\penalty0 (4):\penalty0 292--309, 1997.

\bibitem[Bassett and Sporns(2017)]{bassett2017network}
Danielle~S. Bassett and Olaf Sporns.
\newblock Network neuroscience.
\newblock \emph{Nature Neuroscience}, 20\penalty0 (3):\penalty0 353--364, 2017.
\newblock \doi{10.1038/nn.4502}.

\bibitem[Bechara et~al.(2000)Bechara, Damasio, Damasio, and Lee]{bechara2000emotion}
Antoine Bechara, Hanna Damasio, Antonio~R. Damasio, and Gary~P. Lee.
\newblock Emotion, decision making and the orbitofrontal cortex.
\newblock \emph{Cerebral Cortex}, 10\penalty0 (3):\penalty0 295--307, 2000.
\newblock \doi{10.1093/cercor/10.3.295}.

\bibitem[Blair(2005)]{blair2005neurocognitive}
R.~James~R. Blair.
\newblock Applying a cognitive neuroscience perspective to the disorder of psychopathy.
\newblock \emph{Development and Psychopathology}, 17\penalty0 (3):\penalty0 865--891, 2005.
\newblock \doi{10.1017/S0954579405050418}.

\bibitem[Blair(2007)]{blair2007amygdala_vmPFC}
R.~James~R. Blair.
\newblock The amygdala and ventromedial prefrontal cortex: functional contributions and dysfunction in psychopathy.
\newblock \emph{Philosophical Transactions of the Royal Society B: Biological Sciences}, 363\penalty0 (1503):\penalty0 2557--2565, 2007.
\newblock \doi{10.1098/rstb.2008.0027}.

\bibitem[Boly et~al.(2017)Boly, Massimini, Tsuchiya, Postle, Koch, and Tononi]{boly2017consciousness}
Mélanie Boly, Marcello Massimini, Naotsugu Tsuchiya, Bradley~R. Postle, Christof Koch, and Giulio Tononi.
\newblock Are the neural correlates of consciousness in the front or in the back of the cerebral cortex? clinical and neuroimaging evidence.
\newblock \emph{Journal of Neuroscience}, 37\penalty0 (40):\penalty0 9603--9613, 2017.
\newblock \doi{10.1523/JNEUROSCI.3218-16.2017}.

\bibitem[Brand et~al.(2009)Brand, Eggers, Reinhold, Fujiwara, Kessler, Heiss, and Markowitsch]{brand2009dissociative}
Matthias Brand, Carsten Eggers, Nadine Reinhold, Esther Fujiwara, Josef Kessler, Wolf-Dieter Heiss, and Hans~J. Markowitsch.
\newblock Functional brain imaging in 14 patients with dissociative amnesia reveals right inferolateral prefrontal hypometabolism.
\newblock \emph{Psychiatry Research: Neuroimaging}, 174\penalty0 (1):\penalty0 32--39, 2009.
\newblock \doi{10.1016/j.pscychresns.2009.03.008}.

\bibitem[Buckner et~al.(2008)Buckner, Andrews-Hanna, and Schacter]{buckner2008brain}
Randy~L Buckner, Jessica~R Andrews-Hanna, and Daniel~L Schacter.
\newblock The brain's default network: anatomy, function, and relevance to disease.
\newblock \emph{Annals of the New York Academy of Sciences}, 1124\penalty0 (1):\penalty0 1--38, 2008.
\newblock \doi{10.1196/annals.1440.011}.

\bibitem[Cahill et~al.(2003)Cahill, Gorski, and Le]{cahill2003enhanced}
Larry Cahill, Lukasz Gorski, and Kimberly Le.
\newblock Enhanced human memory consolidation with post-learning stress: Interaction with the degree of arousal at encoding.
\newblock \emph{Learning \& Memory}, 10\penalty0 (4):\penalty0 270--274, 2003.
\newblock \doi{10.1101/lm.62403}.

\bibitem[Casali et~al.(2013)Casali, Gosseries, Rosanova, Boly, Sarasso, Casali, Casarotto, Bruno, Laureys, and Massimini]{casali2013theoretically}
Adenauer~G. Casali, Olivia Gosseries, Mario Rosanova, Mélanie Boly, Simone Sarasso, Kryger~M. Casali, Silvia Casarotto, Marie-Aur{\'e}lie Bruno, Steven Laureys, and Marcello Massimini.
\newblock A theoretically based index of consciousness independent of sensory processing and behavior.
\newblock \emph{Science Translational Medicine}, 5\penalty0 (198):\penalty0 198ra105, 2013.
\newblock \doi{10.1126/scitranslmed.3006294}.

\bibitem[Chalmers(1995)]{chalmers1995facing}
David~J. Chalmers.
\newblock Facing up to the problem of consciousness.
\newblock \emph{Journal of Consciousness Studies}, 2\penalty0 (3):\penalty0 200--219, 1995.

\bibitem[Chrisley(2009)]{chrisley2009synthetic}
Ron Chrisley.
\newblock Synthetic phenomenology.
\newblock \emph{International Journal of Machine Consciousness}, 1\penalty0 (1):\penalty0 53--70, 2009.
\newblock \doi{10.1142/S1793843009000074}.

\bibitem[Conway and Pleydell‑Pearce(2000)]{conway2000sms}
Martin~A. Conway and Christopher~W. Pleydell‑Pearce.
\newblock The construction of autobiographical memories in the self‐memory system.
\newblock \emph{Psychological Review}, 107\penalty0 (2):\penalty0 261--288, 2000.
\newblock \doi{10.1037/0033-295x.107.2.261}.

\bibitem[Corbetta and Shulman(2002)]{corbetta2002control}
Maurizio Corbetta and Gordon~L. Shulman.
\newblock Control of goal-directed and stimulus-driven attention in the brain.
\newblock \emph{Nature Reviews Neuroscience}, 3\penalty0 (3):\penalty0 201--215, 2002.
\newblock \doi{10.1038/nrn755}.

\bibitem[Davisson and Hoyle(2017)]{davisson2017social}
Erin~K. Davisson and Rick~H. Hoyle.
\newblock The social-psychological perspective on self-regulation.
\newblock In T.~Egner, editor, \emph{The Wiley Handbook of Cognitive Control}, pages 469--486. Wiley-Blackwell, 2017.
\newblock \doi{10.1002/9781118920497.ch25}.

\bibitem[de~Waal(2016)]{dewaal2016mama}
Frans B.~M. de~Waal.
\newblock \emph{Mama's Last Hug: Animal Emotions and What They Tell Us about Ourselves}.
\newblock W. W. Norton \& Company, 2016.
\newblock ISBN 9780393635065.

\bibitem[Dehaene and Changeux(2011)]{dehaene2011consciousness}
Stanislas Dehaene and Jean-Pierre Changeux.
\newblock Experimental and theoretical approaches to conscious processing.
\newblock \emph{Neuron}, 70\penalty0 (2):\penalty0 200--227, 2011.
\newblock \doi{10.1016/j.neuron.2011.03.018}.

\bibitem[Dolcos et~al.(2004)Dolcos, LaBar, and Cabeza]{dolcos2004interaction}
Florin Dolcos, Kevin~S LaBar, and Roberto Cabeza.
\newblock Interaction between the amygdala and the medial temporal lobe memory system predicts better memory for emotional events.
\newblock \emph{Neuron}, 42\penalty0 (5):\penalty0 855--863, 2004.
\newblock \doi{10.1016/S0896-6273(04)00289-2}.

\bibitem[Dunn et~al.(2006)Dunn, Dalgleish, and Lawrence]{dunn2006somatic}
Barnaby~D. Dunn, Tim Dalgleish, and Andrew~D. Lawrence.
\newblock The somatic marker hypothesis: A critical evaluation.
\newblock \emph{Neuroscience \& Biobehavioral Reviews}, 30\penalty0 (2):\penalty0 239--271, 2006.
\newblock \doi{10.1016/j.neubiorev.2005.07.001}.

\bibitem[Edelman and Tononi(2000)]{edelman2000universe}
Gerald~M. Edelman and Giulio Tononi.
\newblock \emph{A Universe of Consciousness: How Matter Becomes Imagination}.
\newblock Basic Books, New York, 2000.
\newblock ISBN 9780465013760.

\bibitem[Engel et~al.(2001)Engel, Fries, and Singer]{engel2001dynamic}
Andreas~K. Engel, Pascal Fries, and Wolf Singer.
\newblock Dynamic predictions: oscillations and synchrony in top–down processing.
\newblock \emph{Nature Reviews Neuroscience}, 2\penalty0 (10):\penalty0 704--716, 2001.
\newblock \doi{10.1038/35094565}.

\bibitem[Fodor(1983)]{fodor1983modularity}
Jerry~A. Fodor.
\newblock \emph{The Modularity of Mind: An Essay on Faculty Psychology}.
\newblock MIT Press, Cambridge, MA, 1983.
\newblock ISBN 9780262315920.
\newblock \doi{10.7551/mitpress/4737.001.0001}.

\bibitem[Franklin et~al.(2006)Franklin, Ramamurthy, and D'Mello]{franklin2006lida}
Stan Franklin, Uma Ramamurthy, and Sidney D'Mello.
\newblock The lida architecture: Adding new modes of learning to an intelligent, autonomous, software agent.
\newblock \emph{Integrated Design and Process Technology}, 6\penalty0 (1):\penalty0 1--8, 2006.

\bibitem[Friston and Kiebel(2009)]{friston2009free}
Karl Friston and Stefan Kiebel.
\newblock Predictive coding under the free-energy principle.
\newblock \emph{Philosophical Transactions of the Royal Society B: Biological Sciences}, 364\penalty0 (1521):\penalty0 1211--1221, 2009.
\newblock \doi{10.1098/rstb.2008.0300}.

\bibitem[Frith and Frith(2006)]{frith2006social}
Chris~D. Frith and Uta Frith.
\newblock The neural basis of mentalizing.
\newblock \emph{Neuron}, 50\penalty0 (4):\penalty0 531--534, 2006.
\newblock \doi{10.1016/j.neuron.2006.05.001}.

\bibitem[Gamez(2018)]{gamez2018human}
David Gamez.
\newblock \emph{Human and Machine Consciousness}.
\newblock Open Book Publishers, Cambridge, UK, 2018.
\newblock ISBN 9781783746843.
\newblock \doi{10.11647/OBP.0107}.

\bibitem[Gao et~al.(2009)Gao, Glenn, Schug, Yang, and Raine]{gao2009neurodevelopmental}
Yu~Gao, Andrea~L. Glenn, Robert~A. Schug, Yaling Yang, and Adrian Raine.
\newblock The neurobiology of psychopathy: a neurodevelopmental perspective.
\newblock \emph{Canadian Journal of Psychiatry}, 54\penalty0 (12):\penalty0 813--823, 2009.
\newblock \doi{10.1177/070674370905401204}.

\bibitem[Gazzaniga(2011)]{gazzaniga2011who}
Michael~S. Gazzaniga.
\newblock \emph{Who’s in Charge? Free Will and the Science of the Brain}.
\newblock HarperCollins, New York, 2011.
\newblock ISBN 9780061906107.

\bibitem[Ginsburg and Jablonka(2019)]{ginsburg2019evolution}
Simona Ginsburg and Eva Jablonka.
\newblock \emph{The Evolution of the Sensitive Soul: Learning and the Origins of Consciousness}.
\newblock MIT Press, Cambridge, MA, 2019.
\newblock ISBN 9780262039307.

\bibitem[Goldner‑Vukov and Moore(2010)]{goldner2010malignant}
Mila Goldner‑Vukov and Laurie~Jo Moore.
\newblock Malignant narcissism: From fairy tales to harsh reality.
\newblock \emph{Psychiatria Danubina}, 22\penalty0 (3):\penalty0 392--405, 2010.

\bibitem[Goodwin and Jamison(2007)]{goodwin2007manic}
Frederick~K. Goodwin and Kay~Redfield Jamison.
\newblock \emph{Manic-Depressive Illness: Bipolar Disorders and Recurrent Depression}.
\newblock Oxford University Press, 2nd edition, 2007.
\newblock ISBN 9780195135794.

\bibitem[Graves et~al.(2016)Graves, Wayne, and Danihelka]{graves2016hybrid}
Alex Graves, Greg Wayne, and Ivo Danihelka.
\newblock Hybrid computing using a neural network with dynamic external memory.
\newblock \emph{Nature}, 538\penalty0 (7626):\penalty0 471--476, 2016.
\newblock \doi{10.1038/nature20101}.

\bibitem[Graziano(2017)]{graziano2019attention}
Michael~S.A. Graziano.
\newblock The attention schema theory: A foundation for engineering artificial consciousness.
\newblock \emph{Frontiers in Robotics and AI}, 4, 2017.
\newblock \doi{10.3389/frobt.2017.00060}.

\bibitem[Greene et~al.(2001)Greene, Sommerville, Nystrom, Darley, and Cohen]{greene2001fmri}
Joshua~D Greene, R.~Brian Sommerville, Leigh~E Nystrom, John~M Darley, and Jonathan~D Cohen.
\newblock An fmri investigation of emotional engagement in moral judgment.
\newblock \emph{Science}, 293\penalty0 (5537):\penalty0 2105--2108, 2001.
\newblock \doi{10.1126/science.1062872}.

\bibitem[Guldin and Leget(2023)]{guldin2023integrated}
Mai‑Britt Guldin and Carlo J.~W. Leget.
\newblock The integrated process model of loss and grief – an interprofessional understanding.
\newblock \emph{Death Studies}, page advance online publication, 2023.
\newblock \doi{10.1080/07481187.2023.2272960}.

\bibitem[Hao et~al.(2021)Hao, Zhang, Liu, Wang, Qiao, Liu, Wang, Wang, and Zhang]{hao2021emotion}
Ying Hao, Yiran Zhang, Xiaoxia Liu, Kai Wang, Ying Qiao, Qiang Liu, Ke~Wang, Yonghui Wang, and Qinglin Zhang.
\newblock Neural responses during emotion transitions and regulation: The role of the ventral anterior cingulate cortex and ventromedial prefrontal cortex.
\newblock \emph{Frontiers in Psychology}, 12:\penalty0 666284, 2021.
\newblock \doi{10.3389/fpsyg.2021.666284}.

\bibitem[Hare(1999)]{hare1999without}
Robert~D. Hare.
\newblock \emph{Without Conscience: The Disturbing World of the Psychopaths Among Us}.
\newblock Guilford Press, New York, 1999.
\newblock ISBN 9781572304512.

\bibitem[Jauk and Kanske(2021)]{jauk2021neuroscience}
Emanuel Jauk and Philipp Kanske.
\newblock Can neuroscience help to understand narcissism? a systematic review of an emerging field.
\newblock \emph{Personality Neuroscience}, 4:\penalty0 e3, 2021.
\newblock \doi{10.1017/pen.2021.1}.

\bibitem[Ji et~al.(2016)Ji, Burnett~Heyes, MacLeod, and Holmes]{ji2016emotional}
Julie~L. Ji, Stephanie Burnett~Heyes, Colin MacLeod, and Emily~A. Holmes.
\newblock Emotional mental imagery as simulation of reality: Fear and beyond — a tribute to peter lang.
\newblock \emph{Behavior Therapy}, 47\penalty0 (5):\penalty0 702--719, 2016.
\newblock \doi{10.1016/j.beth.2015.11.004}.

\bibitem[Kahneman(2011)]{kahneman2011thinking}
Daniel Kahneman.
\newblock \emph{Thinking, Fast and Slow}.
\newblock Farrar, Straus and Giroux, New York, 2011.
\newblock ISBN 9780374275631.

\bibitem[Kanwisher(2010)]{kanwisher2010functional}
Nancy Kanwisher.
\newblock Functional specificity in the human brain: A window into the functional architecture of the mind.
\newblock \emph{Proceedings of the National Academy of Sciences}, 107\penalty0 (25):\penalty0 11163--11170, 2010.
\newblock \doi{10.1073/pnas.1005062107}.

\bibitem[Kernberg(1984)]{kernberg1984severe}
Otto~F. Kernberg.
\newblock \emph{Severe Personality Disorders: Psychotherapeutic Strategies}.
\newblock Yale University Press, New Haven, CT, 1984.
\newblock ISBN 9780300053494.

\bibitem[Kiehl(2006)]{kiehl2006psychopathology}
Kent~A. Kiehl.
\newblock A cognitive neuroscience perspective on psychopathy: Evidence for paralimbic system dysfunction.
\newblock \emph{Psychiatry Research}, 142\penalty0 (2-3):\penalty0 107--128, 2006.
\newblock \doi{10.1016/j.psychres.2005.09.013}.

\bibitem[King and Hicks(2021)]{king2021meaning}
Laura~A. King and Joshua~A. Hicks.
\newblock The science of meaning in life.
\newblock \emph{Annual Review of Psychology}, 72:\penalty0 561--584, 2021.
\newblock \doi{10.1146/annurev-psych-072420-122921}.

\bibitem[Koenigs et~al.(2011)Koenigs, Baskin-Sommers, Zeier, and Newman]{koenigs2010neural}
Michael Koenigs, Arielle Baskin-Sommers, Joshua Zeier, and Joseph~P. Newman.
\newblock Investigating the neural correlates of psychopathy: a critical review.
\newblock \emph{Molecular Psychiatry}, 16:\penalty0 792--799, 2011.
\newblock \doi{10.1038/mp.2010.124}.

\bibitem[Lamme(2006)]{lamme2006towards}
Victor A.~F. Lamme.
\newblock Towards a true neural stance on consciousness.
\newblock \emph{Trends in Cognitive Sciences}, 10\penalty0 (11):\penalty0 494--501, 2006.
\newblock \doi{10.1016/j.tics.2006.09.001}.

\bibitem[Lau and Rosenthal(2011)]{lau2011hot}
Hakwan Lau and David Rosenthal.
\newblock Empirical support for higher-order theories of consciousness.
\newblock \emph{Trends in Cognitive Sciences}, 15\penalty0 (8):\penalty0 365--373, 2011.
\newblock \doi{10.1016/j.tics.2011.06.013}.

\bibitem[Libet(1985)]{libet1985unconscious}
Benjamin Libet.
\newblock Unconscious cerebral initiative and the role of conscious will in voluntary action.
\newblock \emph{Behavioral and Brain Sciences}, 8\penalty0 (4):\penalty0 529--539, 1985.
\newblock \doi{10.1017/S0140525X00044903}.

\bibitem[Marino et~al.(2007)Marino, Connor, Fordyce, Herman, Hof, Lefebvre, Lusseau, McCowan, Nimchinsky, Pack, Rendell, Reidenberg, Reiss, Uhen, Van~der Gucht, and Whitehead]{marino2007cetaceans}
Lori Marino, Richard~C. Connor, R.~Ewan Fordyce, Louis~M. Herman, Patrick~R. Hof, Louis Lefebvre, David Lusseau, Brenda McCowan, Esther~A. Nimchinsky, Adam~A. Pack, Luke Rendell, Joy~S. Reidenberg, Diana Reiss, Mark~D. Uhen, Estel Van~der Gucht, and Hal Whitehead.
\newblock Cetaceans have complex brains for complex cognition.
\newblock \emph{PLoS Biology}, 5\penalty0 (5):\penalty0 e139, 2007.
\newblock \doi{10.1371/journal.pbio.0050139}.

\bibitem[Mashour et~al.(2020)Mashour, Roelfsema, Changeux, and Dehaene]{mashour2020consciousness}
George~A. Mashour, Pieter~R. Roelfsema, Jean-Pierre Changeux, and Stanislas Dehaene.
\newblock Conscious processing and the global neuronal workspace hypothesis.
\newblock \emph{Neuron}, 105\penalty0 (5):\penalty0 776--798, 2020.
\newblock \doi{10.1016/j.neuron.2020.01.026}.

\bibitem[Menon(2015)]{menon2015salience}
Vinod Menon.
\newblock Salience network.
\newblock \emph{Brain Mapping: An Encyclopedic Reference}, 2:\penalty0 597--611, 2015.
\newblock \doi{10.1016/B978-0-12-397025-1.00052-X}.

\bibitem[Merker(2007)]{merker2007consciousness}
Bjorn Merker.
\newblock Consciousness without a cerebral cortex: A challenge for neuroscience and medicine.
\newblock \emph{Behavioral and Brain Sciences}, 30\penalty0 (1):\penalty0 63--81, 2007.
\newblock \doi{10.1017/S0140525X07000891}.

\bibitem[Miconi et~al.(2018)Miconi, Clune, and Stanley]{miconi2018differentiable}
Thomas Miconi, Jeff Clune, and Kenneth~O. Stanley.
\newblock Differentiable plasticity: Training plastic neural networks with backpropagation.
\newblock \emph{PLoS One}, 13\penalty0 (2):\penalty0 e0192828, 2018.
\newblock \doi{10.1371/journal.pone.0192828}.

\bibitem[Miller and Cohen(2001)]{miller2001integrative}
Earl~K. Miller and Jonathan~D. Cohen.
\newblock An integrative theory of prefrontal cortex function.
\newblock \emph{Annual Review of Neuroscience}, 24:\penalty0 167--202, 2001.
\newblock \doi{10.1146/annurev.neuro.24.1.167}.

\bibitem[Modesti et~al.(2022)Modesti, Rapisarda, Capriotti, and Del~Casale]{modesti2022functional}
Martina~Nicole Modesti, Ludovica Rapisarda, Gabriela Capriotti, and Antonio Del~Casale.
\newblock Functional neuroimaging in dissociative disorders: A systematic review.
\newblock \emph{Journal of Personalized Medicine}, 12\penalty0 (9):\penalty0 1405, 2022.
\newblock \doi{10.3390/jpm12091405}.

\bibitem[Moll et~al.(2005)Moll, Zahn, de~Oliveira-Souza, Krueger, and Grafman]{moll2005moral}
Jorge Moll, Roland Zahn, Ricardo de~Oliveira-Souza, Frank Krueger, and Jordan Grafman.
\newblock The neural basis of human moral cognition.
\newblock \emph{Nature Reviews Neuroscience}, 6\penalty0 (10):\penalty0 799--809, 2005.
\newblock \doi{10.1038/nrn1768}.

\bibitem[Montijn et~al.(2021)Montijn, Van~den Hout, and Engelhard]{montijn2021emotion}
N.D. Montijn, M.A. Van~den Hout, and I.M. Engelhard.
\newblock Emotion improves memory for imagined future events.
\newblock \emph{Psychological Science}, 32\penalty0 (2):\penalty0 227--237, 2021.
\newblock \doi{10.1177/0956797620972491}.

\bibitem[Morin(2005)]{morin2005inner}
Alain Morin.
\newblock Possible links between self-awareness and inner speech: Theoretical background, underlying mechanisms, and empirical evidence.
\newblock \emph{Journal of Consciousness Studies}, 12\penalty0 (4-5):\penalty0 115--134, 2005.

\bibitem[Moroz(2009)]{moroz2009independent}
Leonid~L. Moroz.
\newblock On the independent origins of complex brains and neurons.
\newblock \emph{Brain, Behavior and Evolution}, 74\penalty0 (3):\penalty0 177--190, 2009.
\newblock \doi{10.1159/000258665}.

\bibitem[Motzkin et~al.(2011)Motzkin, Newman, Kiehl, and Koenigs]{motzkin2011reduced}
Julian~C. Motzkin, Joseph~P. Newman, Kent~A. Kiehl, and Michael Koenigs.
\newblock Reduced prefrontal connectivity in psychopathy.
\newblock \emph{Journal of Neuroscience}, 31\penalty0 (48):\penalty0 17348--17357, 2011.
\newblock \doi{10.1523/JNEUROSCI.4215-11.2011}.

\bibitem[Noonan et~al.(2018)Noonan, Mars, Sallet, Dunbar, and Fellows]{noonan2018socialnetworks}
M.~P. Noonan, R.~B. Mars, J.~Sallet, R.~I.~M. Dunbar, and L.~K. Fellows.
\newblock The structural and functional brain networks that support human social networks.
\newblock \emph{Behavioural Brain Research}, 355:\penalty0 12--23, 2018.
\newblock \doi{10.1016/j.bbr.2018.02.019}.

\bibitem[Oizumi et~al.(2014)Oizumi, Albantakis, and Tononi]{oizumi2014phenomenology}
Masafumi Oizumi, Larissa Albantakis, and Giulio Tononi.
\newblock From the phenomenology to the mechanisms of consciousness: Integrated information theory 3.0.
\newblock \emph{PLoS Computational Biology}, 10\penalty0 (5):\penalty0 e1003588, 2014.
\newblock \doi{10.1371/journal.pcbi.1003588}.

\bibitem[Patrick(2018)]{patrick2018handbook}
Christopher~J. Patrick.
\newblock \emph{Handbook of Psychopathy}.
\newblock Guilford Press, New York, 2018.
\newblock ISBN 9781462541232.

\bibitem[Peters and B{\"u}chel(2010)]{peters2010future}
Jan Peters and Christian B{\"u}chel.
\newblock Episodic future thinking reduces reward delay discounting through an enhancement of prefrontal-mediotemporal interactions.
\newblock \emph{Neuron}, 66\penalty0 (1):\penalty0 138--148, 2010.
\newblock \doi{10.1016/j.neuron.2010.03.026}.

\bibitem[Phan et~al.(2002)Phan, Wager, Taylor, and Liberzon]{phan2002functional}
K.~Luan Phan, Tor~D. Wager, Stephan~F. Taylor, and Israel Liberzon.
\newblock Functional neuroanatomy of emotion: a meta-analysis of emotion activation studies in pet and fmri.
\newblock \emph{NeuroImage}, 16\penalty0 (2):\penalty0 331--348, 2002.
\newblock \doi{10.1006/nimg.2002.1087}.

\bibitem[Plotnik et~al.(2006)Plotnik, de~Waal, and Reiss]{plotnik2006mirror}
Joshua~M Plotnik, Frans B~M de~Waal, and Diana Reiss.
\newblock Self-recognition in an asian elephant.
\newblock \emph{Proceedings of the National Academy of Sciences}, 103\penalty0 (45):\penalty0 17053--17057, 2006.
\newblock \doi{10.1073/pnas.0608062103}.

\bibitem[Redinbaugh et~al.(2020)Redinbaugh, Phillips, Kambi, Mohanta, Andryk, Dooley, Afrasiabi, Raz, and Saalmann]{redinbaugh2020thalamus}
Meg~J. Redinbaugh, Jennifer~M. Phillips, Niranjan~A. Kambi, Sudeep Mohanta, Sonia Andryk, Gwendolyn~L. Dooley, Mojtaba Afrasiabi, Avraham Raz, and Yuri~B. Saalmann.
\newblock Thalamus modulates consciousness via layer-specific control of cortex.
\newblock \emph{Neuron}, 106\penalty0 (1):\penalty0 66--75.e12, 2020.
\newblock \doi{10.1016/j.neuron.2020.01.005}.

\bibitem[Reinders et~al.(2012)Reinders, Willemsen, Vos, den Boer, and Nijenhuis]{reinders2012fact}
Antje A. T.~S. Reinders, Antoon T.~M. Willemsen, Herry P.~J. Vos, Johan~A. den Boer, and Ellert R.~S. Nijenhuis.
\newblock Fact or factitious? a psychobiological study of authentic and simulated dissociative identity states.
\newblock \emph{PLoS ONE}, 7\penalty0 (6):\penalty0 e39279, 2012.
\newblock \doi{10.1371/journal.pone.0039279}.

\bibitem[Rhodewalt and Eddings(2002)]{rhodewalt2002narcissus}
Frederick Rhodewalt and Stacy~K. Eddings.
\newblock Narcissus reflects: Memory distortion in response to ego-relevant feedback among high- and low-narcissistic men.
\newblock \emph{Journal of Research in Personality}, 36\penalty0 (2):\penalty0 97--116, 2002.
\newblock \doi{10.1006/jrpe.2002.2342}.

\bibitem[Robjant and Fazel(2010)]{robjant2010emerging}
Katy Robjant and Mina Fazel.
\newblock The emerging evidence for narrative exposure therapy.
\newblock \emph{Clinical Psychology Review}, 30\penalty0 (8):\penalty0 1030--1039, 2010.
\newblock \doi{10.1016/j.cpr.2010.07.004}.

\bibitem[Roozendaal et~al.(2009)Roozendaal, McEwen, and Chattarji]{roozendaal2009amygdala}
Benno Roozendaal, Bruce~S. McEwen, and Sumantra Chattarji.
\newblock Stress, memory and the amygdala.
\newblock \emph{Nature Reviews Neuroscience}, 10\penalty0 (6):\penalty0 423--433, 2009.
\newblock \doi{10.1038/nrn2651}.

\bibitem[Rosenthal(2005)]{rosenthal2005consciousness}
David~M. Rosenthal.
\newblock \emph{Consciousness and Mind}.
\newblock Oxford University Press, 2005.
\newblock ISBN 978-0198236962.

\bibitem[Schacter and Addis(2007)]{schacter2007constructive}
Daniel~L. Schacter and Donna~Rose Addis.
\newblock The cognitive neuroscience of constructive memory: remembering the past and imagining the future.
\newblock \emph{Philosophical Transactions of the Royal Society B: Biological Sciences}, 362\penalty0 (1481):\penalty0 773--786, 2007.
\newblock \doi{10.1098/rstb.2007.2087}.

\bibitem[Schiff et~al.(2007)Schiff, Giacino, Kalmar, Victor, Baker, Gerber, Fritz, Eisenberg, Biondi, O’Connor, Kobylarz, Farris, Machado, McCagg, Plum, Fins, and Rezai]{schiff2007behavioral}
Nicholas~D. Schiff, Joseph~T. Giacino, Kathleen Kalmar, Jonathan~D. Victor, Kimberly Baker, Martin Gerber, Barbara Fritz, Bruce Eisenberg, Tim Biondi, John O’Connor, Evan~J. Kobylarz, Susan Farris, Andre Machado, Conor McCagg, Fred Plum, Joseph~J. Fins, and Ali~R. Rezai.
\newblock Behavioral improvements with thalamic stimulation after severe traumatic brain injury.
\newblock \emph{Nature}, 448\penalty0 (7153):\penalty0 600--603, 2007.
\newblock \doi{10.1038/nature06041}.

\bibitem[Schmidt et~al.(2023)Schmidt, Pfarr, Meller, Evermann, and Nenadi{\'c}]{schmidt2023structural}
Lisa Schmidt, Julia-Katharina Pfarr, Tina Meller, Uwe Evermann, and Igor Nenadi{\'c}.
\newblock Structural connectivity of grandiose versus vulnerable narcissism as models of social dominance and subordination.
\newblock \emph{Scientific Reports}, 13\penalty0 (1):\penalty0 16098, 2023.
\newblock \doi{10.1038/s41598-023-41098-1}.

\bibitem[Schurger et~al.(2012)Schurger, Sitt, and Dehaene]{schurger2012accumulator}
Aaron Schurger, Jacobo~D. Sitt, and Stanislas Dehaene.
\newblock An accumulator model for spontaneous neural activity prior to self-initiated movement.
\newblock \emph{Proceedings of the National Academy of Sciences}, 109\penalty0 (42):\penalty0 E2904--E2913, 2012.
\newblock \doi{10.1073/pnas.1210467109}.

\bibitem[Sedikides et~al.(2023)Sedikides, Hong, and Wildschut]{sedikides2023self}
Constantine Sedikides, Emily~K. Hong, and Tim Wildschut.
\newblock Self‑continuity: An integrative framework and new directions.
\newblock \emph{Annual Review of Psychology}, 74:\penalty0 333--361, 2023.
\newblock \doi{10.1146/annurev-psych-032420-032236}.

\bibitem[Seeley et~al.(2007)Seeley, Menon, Schatzberg, Keller, Glover, Kenna, Reiss, and Greicius]{seeley2007dissociable}
William~W. Seeley, Vinod Menon, Alan~F. Schatzberg, Jonathan Keller, Gary~H. Glover, Heather Kenna, Allan~L. Reiss, and Michael~D. Greicius.
\newblock Dissociable intrinsic connectivity networks for salience processing and executive control.
\newblock \emph{Journal of Neuroscience}, 27\penalty0 (9):\penalty0 2349--2356, 2007.
\newblock \doi{10.1523/JNEUROSCI.5587-06.2007}.

\bibitem[Sen and Pakyürek(2023)]{sen2023mediating}
Gamze Sen and Gün Pakyürek.
\newblock The mediating role of autobiographical memory in the relationship between narcissism and rejection sensitivity.
\newblock \emph{Psychological Reports}, 126\penalty0 (2):\penalty0 856--876, 2023.
\newblock \doi{10.1177/00332941211061076}.

\bibitem[Somma et~al.(2024)Somma, Borroni, and Krueger]{somma2024assessing}
Alberto Somma, Serena Borroni, and Robert~F. Krueger.
\newblock Assessing malignant narcissism and its associations with maladaptive personality traits: Preliminary evidence from the pid-5 model.
\newblock \emph{Mediterranean Journal of Clinical Psychology}, 12\penalty0 (1):\penalty0 1--17, 2024.
\newblock \doi{10.13129/2282-1619/mjcp-4079}.

\bibitem[Spreng et~al.(2010)Spreng, Sepulcre, Turner, Stevens, and Schacter]{spreng2010default}
R.~Nathan Spreng, Jorge Sepulcre, Glenn~R. Turner, W.~Kyle Stevens, and Daniel~L. Schacter.
\newblock Default network activity, coupled with the frontoparietal control network, supports goal-directed cognition.
\newblock \emph{NeuroImage}, 53\penalty0 (1):\penalty0 303--317, 2010.
\newblock \doi{10.1016/j.neuroimage.2010.06.016}.

\bibitem[Tangney et~al.(2007)Tangney, Stuewig, and Mashek]{tangney2007self}
June~Price Tangney, Jeff Stuewig, and Debra~J. Mashek.
\newblock Self-conscious emotions: The self as a moral guide.
\newblock In Jessica~L. Tracy, Richard~W. Robins, and June~Price Tangney, editors, \emph{The self-conscious emotions: Theory and research}, pages 1--22. Guilford Press, New York, 2007.
\newblock ISBN 9781593850407.

\bibitem[Taïb et~al.(2023)]{taib2023neural}
Sarah Taïb et~al.
\newblock What are the neural correlates of dissociative amnesia? a scoping review of neuroimaging studies.
\newblock \emph{Frontiers in Psychiatry}, 14:\penalty0 1092826, 2023.
\newblock \doi{10.3389/fpsyt.2023.1092826}.

\bibitem[Thome et~al.(2019)Thome, Terpou, McKinnon, and Lanius]{thome2020neural}
Janine Thome, Braeden~A. Terpou, Margaret~C. McKinnon, and Ruth~A. Lanius.
\newblock The neural correlates of trauma-related autobiographical memory in posttraumatic stress disorder: A meta-analysis.
\newblock \emph{Depression and Anxiety}, 2019.
\newblock \doi{10.1002/da.22977}.

\bibitem[Tononi(2004)]{tononi2004information}
Giulio Tononi.
\newblock An information integration theory of consciousness.
\newblock \emph{BMC Neuroscience}, 5\penalty0 (1):\penalty0 42, 2004.
\newblock \doi{10.1186/1471-2202-5-42}.

\bibitem[Tulving(1983)]{tulving1983elements}
Endel Tulving.
\newblock Elements of episodic memory.
\newblock In \emph{Memory and consciousness}, pages 1--45. Oxford University Press, 1983.

\bibitem[VanRullen and Koch(2003)]{vanrullen2003perception}
Rufin VanRullen and Christof Koch.
\newblock Is perception discrete or continuous?
\newblock \emph{Trends in Cognitive Sciences}, 7\penalty0 (5):\penalty0 207--213, 2003.
\newblock \doi{10.1016/S1364-6613(03)00095-0}.

\bibitem[Varela et~al.(2001)Varela, Lachaux, Rodriguez, and Martinerie]{varela2001brainweb}
Francisco~J. Varela, Jean-Philippe Lachaux, Eugenio Rodriguez, and Jacques Martinerie.
\newblock The brainweb: Phase synchronization and large-scale integration.
\newblock \emph{Nature Reviews Neuroscience}, 2\penalty0 (4):\penalty0 229--239, 2001.
\newblock \doi{10.1038/35067550}.

\bibitem[Vaswani et~al.(2017)Vaswani, Shazeer, Parmar, Uszkoreit, Jones, Gomez, Kaiser, and Polosukhin]{vaswani2017attention}
Ashish Vaswani, Noam Shazeer, Niki Parmar, Jakob Uszkoreit, Llion Jones, Aidan~N. Gomez, {\L}ukasz Kaiser, and Illia Polosukhin.
\newblock Attention is all you need.
\newblock In \emph{Advances in Neural Information Processing Systems (NeurIPS)}, volume~30, 2017.
\newblock URL \url{https://papers.nips.cc/paper_files/paper/2017/hash/3f5ee243547dee91fbd053c1c4a845aa-Abstract.html}.

\bibitem[Vincent et~al.(2008)Vincent, Kahn, Snyder, Raichle, and Buckner]{vincent2008frontoparietal}
Justin~L. Vincent, Itamar Kahn, Abraham~Z. Snyder, Marcus~E. Raichle, and Randy~L. Buckner.
\newblock Evidence for a frontoparietal control system revealed by intrinsic functional connectivity.
\newblock \emph{Journal of Neurophysiology}, 100\penalty0 (6):\penalty0 3328--3342, 2008.
\newblock \doi{10.1152/jn.90355.2008}.

\end{thebibliography}

\section*{Figures and Tables}

\newpage

\begin{table}[h!]
\centering
\begin{tabularx}{\textwidth}{|l|l|X|}
\hline
\textbf{Level} &
\textbf{Module} & \textbf{Candidate neural substrates} \\
\hline
Subconscious & Memory module & Hippocampus, entorhinal cortex, parahippocampal gyrus \\
& Behavioral readiness module & Amygdala, hypothalamus, periaqueductal gray, brainstem nuclei \\
& Decision module & Basal ganglia (especially striatum), cerebellum; cortico-subcortical loops involved in action selection and automatized responses \\
\hline
Conscious & 
Filter module & Fronto-parietal attentional networks, intralaminar thalamic nuclei, pulvinar \\
& Narrative module &  Medial prefrontal cortex, precuneus, posterior cingulate cortex (default mode network) \\
& Abstraction module &  Lateral prefrontal cortex, inferior parietal lobule \\
& Evaluation module &  Temporoparietal junction, medial prefrontal cortex, precuneus \\
& Self-evaluation module & Ventromedial prefrontal cortex, anterior cingulate cortex \\
& Integration module & Medial cortical hubs (mPFC, precuneus), fronto-parietal synchrony; possible coordination via the central lateral thalamic nucleus \\
\hline
\end{tabularx}
\caption{\textbf{Tentative mapping between MCT modules and candidate neural substrates.}}
\label{tab:mct_neuro}
\end{table}

\begin{landscape}

\begin{table}[htbp]
\scriptsize
\centering
\begin{adjustbox}{width=\linewidth,center}
\begin{tabular}{|>{\raggedright\arraybackslash}p{3.3cm}|>{\raggedright\arraybackslash}p{2.6cm}|>{\raggedright\arraybackslash}p{2.6cm}|>{\raggedright\arraybackslash}p{3.3cm}|>{\raggedright\arraybackslash}p{2.6cm}|>{\raggedright\arraybackslash}p{2.6cm}|}
\hline
\textbf{Module} & \textbf{Blindsight} & \textbf{Psychopathy} & \textbf{Malignant covert narcissism} & \textbf{Severe dissociation}  & \textbf{Bipolar disorder} \\
\hline
Filter & Impaired (visual signals rejected) & Intact & Intact & Intact & Intact \\
\hline
Narration & Intact (fills perceptual gaps) & Intact; used strategically for dominance & Biased to self-idealization & Fragmented, unstable & Instable, oscillating narratives \\
\hline
Abstraction & Intact & Intact & Partially inhibited, constrained by narrative bias & Partially inhibited  & Amplified or inhibited \\
\hline
Existential submodule & Intact & Impaired (no fear of death) & Atrophied or distorted & Often disconnected & Hyperactive or collapsed \\
\hline
Integration & IIS lacks visual content & Intact & Intact & Fragmented IIS; impaired global coherence &  Intact \\
\hline
Behavioral readiness & Intact & Antipathy hyperactive, empathy absent (sadism possible) & Antipathy hyperactive, empathy absent (manipulation/sadism) & Blunted affect; emotional signals fail to contribute to the construction of IISs & Unstable (euphoria / despair) \\
\hline
Self-evaluation & Intact & Guilt, shame, remorse submodules inactive; pride functional & Guilt submodule distorted to victimization, shame inactive, pride hypertrophied & Partially inhibited; detachment from emotional self-monitoring & Unstable (inflated or crushed self-view) \\
\hline
Evaluation & Intact & Functional, morally unconstrained & Intact & Impaired  & Dysregulated (sensitive or indifferent) \\
\hline
Memory & Intact & Intact & Emotionally distorted, selectively reconstructed & Fragmented or dissociated & Fluctuating recall \\
\hline
\end{tabular}
\vspace{0.5em}
\end{adjustbox}
\caption{\textbf{Modular dysfunctions associated with various clinical pathologies as modeled by MCT}. "Intact" indicates structural and functional preservation of the module.}
\label{tab1}
\end{table}
\end{landscape}

\end{document}